\pdfoutput=1
\documentclass[]{aa}
\usepackage{graphicx}
\usepackage{natbib}
\usepackage{graphicx}
\usepackage{epsfig}
\usepackage{natbib}
\usepackage{textcomp}
\usepackage{amssymb}
\usepackage[figuresright]{rotating}
\newcommand{\micron}{$\mu m$}
\newcommand{\Lx}{$L_{2-10 keV}$}

\newcommand{\xo}{F(2-10 keV)/F(R) }
\newcommand{\miro}{F(24 \micron)/F(R) }
\newcommand{\niro}{F(3.6 \micron)/F(R) }

\newcommand{\perc}{$\%$}

\newcommand{\agn}{Active Galactic Nuclei}
\newcommand{\ergs}{erg s$^{-1}$ }
\def\chandra{{\em Chandra}}
\def\xmm{{\em XMM--Newton}}
\def\spitzer{{\em Spitzer}}
\def\nh{{N$_{\rm H}$}}

\def\XMM{{XMMES1\_}}
\def\R{{\em R}}

\def\ls{{_<\atop^{\sim}}}

\def\cgs{ ${\rm erg~cm}^{-2}~{\rm s}^{-1}$ } 

\begin{document}
\title{The XMM-Newton survey of the ELAIS-S1 field II:  optical identifications and multiwavelength catalogue of X-ray sources
\thanks{Based on observations collected at the 
European Southern Observatory, Prog.ID 073.A-0446(B), 075.A-0428(B) ,076.A-0225(A), 077.A-0800(A) and 078.A-0795(B). 
Based also on observations made with XMM-Newton, an ESA science mission, and with Chandra X-ray Observatory.}}

\author{C. Feruglio\inst{1,2}, F. Fiore\inst{1}, F. La Franca\inst{3}, N. Sacchi\inst{3}, S. Puccetti\inst{4}, A. Comastri\inst{5}, S. Berta\inst{6}, M. Brusa\inst{6}, A. Franceschini\inst{7}, C. Gruppioni\inst{5}, S. Mathur\inst{8}, I. Matute\inst{9}, M. Mignoli\inst{5}, F. Pozzi\inst{10}, C. Vignali\inst{10}, G. Zamorani\inst{5} 
}
{
\institute {INAF~-~Osservatorio Astronomico di Roma 
via Frascati 33, 00040 Monteporzio Catone, Italy
\and
CEA, Irfu, Service d' Astrophysique, Centre de Saclay, F-91191 Gif-sur-Yvette, France\\ 
\email{chiara.feruglio@cea.fr}
\and
Dipartimento di Fisica, Universit\`a Roma Tre, via della Vasca Navale 84, 00146 Roma, Italy
\and
ASI Science Data Center, via G. Galilei, 00044 Frascati, Italy
\and
INAF~-~Osservatorio Astronomico di Bologna, via Ranzani 1, 40127 Bologna, Italy
\and
Max Planck Institut f\"{u}r extraterrestrische Physik, Giessenbachstrasse 1, D-85748 Garching, Germany
\and
Dip. di Astronomia, Universit\'a di Padova, vicolo dell' Osservatorio 3, 35122 Padova, Italy  
\and
Ohio State University, 140 West 18th Avenue, Columbus, OH 43210, USA
\and
INAF~-~Osservatorio Astronomico di Arcetri, Largo E. Fermi 5, 50125 Firenze, Italy
\and
Dipartimento di Astronomia, Universit\'a di Bologna, via Ranzani 1, 40127 Bologna, Italy
}

\date{June 29, 2008}
\abstract{ 
We present the optical identifications and a multi-band catalogue of a sample of 478 X-ray sources detected in the \xmm~  and \chandra~ surveys of the central 0.6 deg$^2$ of the ELAIS-S1 field.
The most likely optical/infrared counterpart of each X-ray source was identified using the chance coincidence probability in the R and IRAC 3.6 \micron~ bands.
This method was complemented by the precise positions obtained through \chandra~ observations.
We were able to associate a counterpart to each X-ray source in the catalogue. 
Approximately 94\% of them are detected in the R band, while
the remaining are blank fields in the optical down to R$\sim$24.5, 
but have a near-infrared counterpart detected by IRAC within 6\arcsec~ from the \xmm~ centroid.
The multi-band catalogue, produced using the positions of the identified optical counterparts, contains  photometry in ten photometric bands, from B to the MIPS 24 \micron~ band.
The spectroscopic follow-up allowed us to determine the redshift and classification for 237 sources ($\sim 50 \%$ of the sample) brighter than $R=24$. 
The spectroscopic redshifts were complemented by reliable photometric redshifts for 68 sources.
We classified 47\% of the sources with spectroscopic redshift as broad-line active galactic nuclei (BL AGNs) with z$=0.1-3.5$, while sources without broad-lines (NOT BL AGNs) are about 46\perc~ of the spectroscopic sample and are found up to $z=2.6$. 
The remaining fraction is represented by extended X-ray sources and stars.
We spectroscopically identified 11 type 2 QSOs among the sources with \xo$>$8, with redshift between 0.9 and 2.6, high 2-10 keV luminosity  (log\Lx$\ge$43.8 \ergs) and hard X-ray colors suggesting large absorbing columns at the rest frame (log$N_H$ up to 23.6 cm$^{-2}$). 
BL AGNs show on average blue optical-to-near-infrared colors, softer X-ray colors and X-ray-to-optical colors typical of optically selected AGNs.  
Conversely, narrow-line sources show redder optical colors, harder X-ray flux ratio and span a wider range of X-ray-to-optical colors.
On average the Spectral Energy Distributions (SEDs) of high-luminosity BL AGNs resemble the power-law typical of unobscured AGNs. The SEDs of NOT BL AGNs are dominated by the galaxy emission in the optical/near-infrared, and show a rise in the mid-infrared which suggests the presence of an obscured active nucleus.
We study the infrared-to-optical colors and near-infrared SEDs to infer the properties of the AGN host galaxies. 
}

\authorrunning {Feruglio}
\titlerunning {The XMM-Newton survey of the ELAIS-S1 field}

\maketitle

\section{Introduction}
One of the primary goals of observational cosmology is the
determination of the census of the \agn~ population in the Universe,
along with their cosmic evolution and the assembly of
super-massive black-holes (SMBHs) in galaxy nuclei.  Several
theoretical and observational results indicate that the assembly of
SMBHs is tightly linked to the evolution of the galaxy bulge
component.  The discovery of SMBHs in the center of most nearby
bulge-dominated galaxies and the correlation existing between the
black-hole mass and the bulge properties (the $M_{BH}-\sigma$
relation, \citet{Gebhardt2000}, \citet{Ferrarese2000}) suggest that
the assembly of bulge masses is tied to the evolution of the accretion
processes in AGNs.  In such scenario, the details of the co-evolution
of black-holes and their host galaxies depend on feedback mechanisms
between the AGNs and their host galaxies (e.g. \citet{Granato2001,Granato2004},
\citet{DiMatteo2005}, \citet{Menci2006}, \citet{Croton2006}).
Luminous AGNs are more efficient in inhibiting the star-formation in
their host galaxies, heating the interstellar matter through winds,
shocks, and high energy radiation, thus making their colors redder.
In this picture,
the AGN phase should precede the phase when a galaxy is caught in a passive phase. 
Indeed, \citet{Pozzi2007} and \citet{Mignoli2004} using Spitzer photometry found that a sample of highly obscured (log$N_H=22.5-23.5$ cm$^{-2}$) QSOs at z=1-2 are hosted by red passive galaxies, suggesting a later stage in their evolution. 
Spectroscopy of type 2 QSOs with the Infrared Spectrograph (IRS) onboard \emph{Spitzer} finds similar results (\citet{Weedman2006}, \citet{Houck2005}). 
This would suggest that the red colors of the IR selected heavily obscured AGNs 
(N$_H \ge 10^{23}$ cm$^{-2}$) may be associated to a passive host galaxy. 
The study of the optical and
infrared colors of AGN host galaxies can therefore put constraints on
AGN feedback mechanisms and on their relative time-scales. This study
is of course easier in the cases where the nuclear light does not
over-shine the stellar light, i.e. in optically obscured AGNs.

Hard X-ray surveys performed by \chandra~ and \xmm~ in 2-10 keV band are
the primary and most efficient tool to detect unobscured and moderately
obscured (N$_H$ up to a few 10$^{23}$ cm$^{-2}$) AGNs up to high
redshift, as they are less affected by dust and gas obscuration,
compared to optical surveys and soft X-ray surveys.  In particular,
deep pencil-beam surveys and shallower, large area surveys have been
recently used to select and study the hard X-ray population, which
dominates the cosmic X-ray background (see \citet{Brandt2005} and references therein).
These surveys are particularly efficient in selecting optically
obscured AGNs, which show X-ray-to-optical colors
(\xo\footnote{$log \xo = log f_X + R/2.5 +5.4176$}, hereafter X/O) larger than 10, and
are usually missed by optical selection (see
e.g. \citet{Fiore2003}, \citet{Cocchia2007}, \citet{Caccianiga2007}).
The majority of these
sources are thought to be obscured QSO at z$\gtrsim$1, but many of
them remain spectroscopically unidentified due to the faintness of
their optical counterparts, which makes them difficult to access
even with 10 m class telescopes.  Several studies suggest that the
amount of obscuration decreases with increasing intrinsic luminosity
(e.g., \citet{LaFranca2005},\citet{Ueda2003}).  
However, even the 2-10 keV selection
becomes highly incomplete in selecting highly obscured and
Compton-thick AGNs, with log\nh$\ge 24 cm^{-2}$.

The \xmm~ survey of ELAIS-S1 is a large-area, medium depth survey, and
therefore is particularly suited to select a large number of luminous
AGNs.  
The ELAIS-S1 field 
covers an area of about 4 deg$^2$ in the southern hemisphere, and
includes the minimum in the Galactic 100 $\mu$m emission in that hemisphere (0.37 MJy/sr, \citet{Schlegel1998}).  
A central 0.6 deg$^2$ contiguous area (centered at 00 34 40.4, -43 28 44.6) has been surveyed with \xmm~ for a total of 400 ks.
The regions with highest XMM sensitivity
($\sim$65\% of the full XMM area) were target of  6 \chandra~
pointings (two of them centered on the same coordinates) with the aim of obtaining precise positions for the
X-ray sources.  The field has a multi-band photometric coverage from
the optical B band to the mid-infrared \spitzer~ bands.  
A spectroscopic follow-up has been performed with VIMOS/VLT, FORS2/VLT
and EFOSC/ESO3.6m for redshift determination and
classification of the sources.

This paper presents the optical identifications and a multi-band
catalogue of the X-ray sample described in \citet{Puccetti2006}, along
with the optical spectroscopy (redshift and source classification) for
a sub-sample of sources, and an analysis of their multiwavelength
properties. In this paper we concentrate on two main issues: 1) the
identifications of a sizable sample of high luminosity, optically
obscured QSOs; 2) the study of the optical and infrared properties
of the host galaxies of the highly obscured AGNs.

The paper is organized as follows: Section 2 presents the
sample, the multi-band photometry and optical spectroscopy.  Section 3
discusses source identification and presents the multi-band catalogue.
Section 4 presents our results, redshifts, source classification and
an analysis of the spectral energy distributions.  In Section 5 we
present our conclusions.  
Magnitudes are given in Vega system unless otherwise stated. 
A $H_0=70$ km s$^{-1}$ Mpc$^{-1}$,
$\Omega_M$=0.3, $\Omega_{\Lambda}=0.7$ cosmology is adopted
throughout.

\section{Observations and data analysis}

\subsection{\xmm~ data}
\xmm~ surveyed the central 0.6 deg$^2$ of the ELAIS-S1 field down to flux limits of 2.5$\times$10$^{-15}$ \cgs in the
2-10 keV band (hard band, H) and $\sim$5.5$\times$10$^{-16}$ \cgs in the 0.5-2 keV band (soft band, S) §\citep{Puccetti2006}. 
A total of 478 sources were detected: 395 in the soft band, and 205 in the hard band.
Among these, \citet{Puccetti2006} identified 7 extended sources, probably groups or clusters.

\subsection{Chandra X-ray data}
\chandra~ observations, whose angular resolution can reach 0.5\arcsec~ on axis, allow us to significantly improve the identification of the optical counterpart.
\chandra~ observations were devised to cover the areas with the highest \xmm~ sensitivity.
The five Chandra pointings have exposure times 32150 s, 37081 s, 40256
s, 23974 s, 29948 s, and are slightly overlapping.  The background in
the non averlapping areas is rather stable ranging from
$1.9\times10^{-7}$ and $2.5\times10^{-7}$ counts/s/pixel or $0.3-0.4$
counts/30 ksec over an area of 2 arcsec radius. 
We are therefore never
background limited for point source detection. Assuming 5 counts for
the faintest detection (which corresponds to a probability of being a
background fluctuation given the above background between $10^{-5}$
and $6\times10^{-5}$) results in a 0.5-7 keV and 0.5-2 keV flux limits
of $\sim1.7\times10^{-15}$ and $\sim6\times10^{-16}$ erg cm$^{-2}$
s$^{-1}$ (for a power law spectrum with $\alpha_E=0.4$).

The five fields were astrometrically corrected by matching the
positions of bright X-ray sources (AGNs) with that of point-like
optical counterparts.  The systematic shifts between the X-ray and the
optical positions were always smaller than a few arcsec.  The residual
systematic shift is $\ls0.1$ arcsec in all five cases.

Source detection was performed on each single event file in the full
0.5-7 keV energy band using the PWDetect code, developed at
INAF~-~Osservatorio Astronomico di Palermo (see \citet{Damiani1997} and \citet{Puccetti2006} for more details on the detection
algorithm).  The significance threshold for each field was chosen in
order to get entries with more than 4-5 counts.

The main technical goal of this paper is the identification of the
counterparts of the XMM X-ray sources. Chandra data are used here only
to help the identification process.  Therefore, the main concern about
the Chandra data analysis is the quality of the positions. Position
uncertainties are proportional to the size of the Chandra point spread
function at the off-axis angle where the source is detected and
proportional to the inverse of the square root of the source counts.
Typical 1 $\sigma$ uncertainties ranges from 0.1 arcsec (limited by
the accuracy of the astrometric correction of the Chandra fields) to a
1-2 arcsec for faint, off-axis sources. This is always better than the
XMM position accuracy. When available, we therefore always used
Chandra positions to help and/or confirm the identification of the
counterparts of the XMM sources.

296 XMM sources are present in the area covered by Chandra
pointings. 239 of these are detected also by Chandra (80\%). 
This is consistent with the Chandra slightly shallower flux limit. 
The Chandra
detections starts to drop strongly below  $\sim 7-8\times10^{-16}$ erg cm$^{-2}$
s$^{-1}$, and
are correspondingly largely incomplete below this flux limit. 
In summary we have rather
accurate Chandra positions for 50\% of the XMM source sample.

\begin{figure}
\centering
\begin{tabular}{c}
\includegraphics[width=8.5cm]{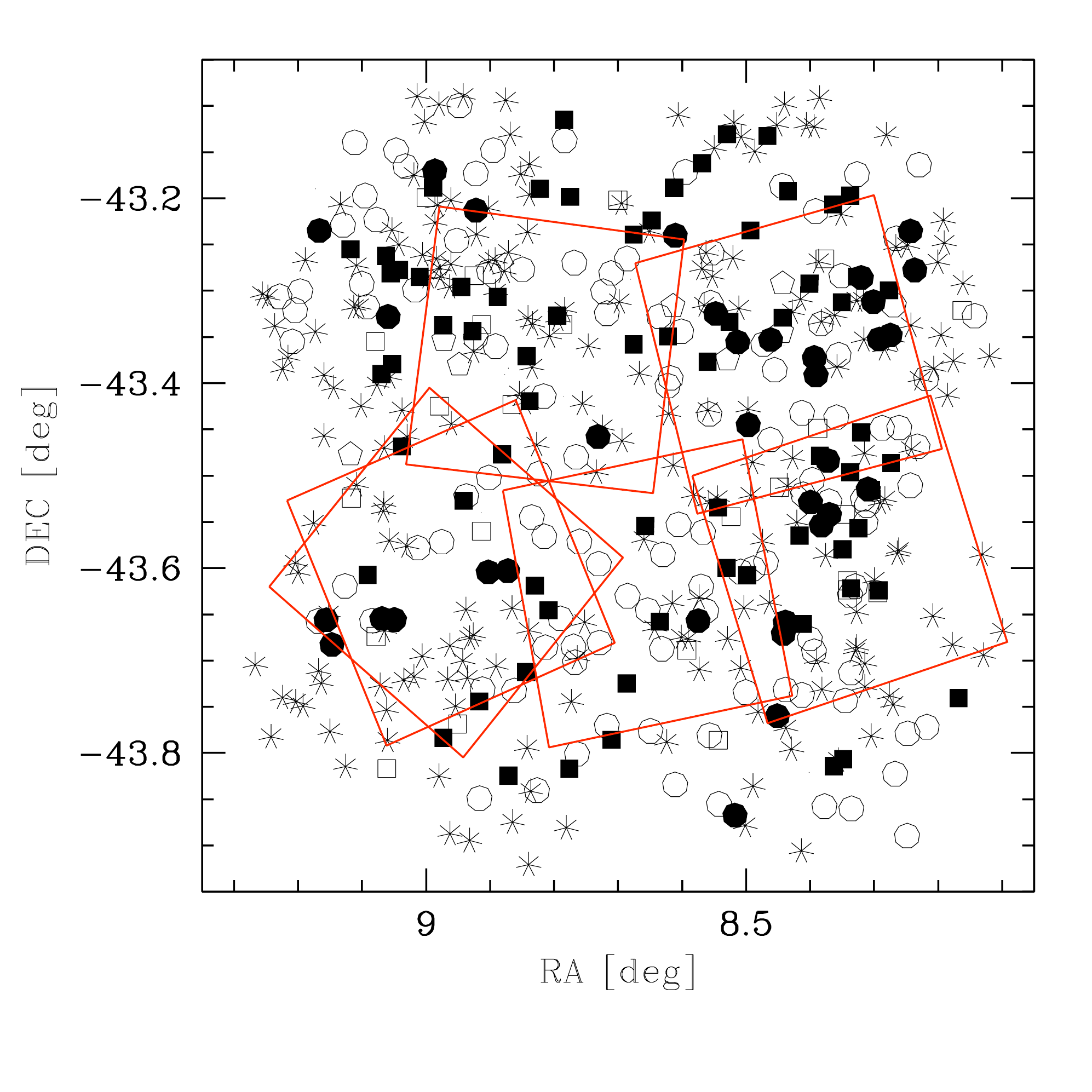}
\end{tabular}
\caption{The distribution of the X-ray sources in the sky.
Open circles = broad-line AGN; filled circles = narrow-line AGN;  filled squares = emission-line galaxies; open squares = early-type galaxies; open pentagons = extended sources; stars = objects without spectroscopic redshift. The large open squares represent the \chandra~ pointings. 
}
\label{radec}
\end{figure}

\subsection{Optical/IR photometry}

ELAIS-S1 is the main southern field of the \emph{Spitzer Wide-area Infra-Red Extragalactic} survey (SWIRE, \citet{Lonsdale2003}, \citet{Lonsdale2004}). 
The  field was observed by Spitzer in December  2004,  with the Infrared Array Camera (IRAC) in the near-Infrared and  the Multiple Imaging Photometer (MIPS) in the mid-Infrared, down to flux limits 4.1 $\mu$Jy at 3.6 $\mu m$ and 0.4 mJy at 24 $\mu m$ (5$\sigma$).
The absolute flux uncertainty is about 10\% , as described in the SWIRE data release \citep{Surace2004}. 

The ELAIS-S1 field is covered also with BVR band photometry from  the ESO-Spitzer Imaging Extragalactic WFI survey  (ESIS, \citet{Berta2006}), performed with the ESO Wide Field Imager. 

The field was observed between July 2004 and October 2005 with VIMOS/VLT \citep{LeFevre2003} to obtain photometry in the R band down to R $\approx$ 24.5, with integration time of 180 seconds.
The pointings were devised to obtain the most complete coverage possible of the area covered by XMM observations, filling the gaps between the VIMOS quadrants.

The XMM ELAIS-S1 field was observed with SOFI/NTT  in J and Ks bands, down to J=21 and K$_s$=19.5 (Vega) within the framework of the ESO Large Program 170.A-0143.

The R and K$_s$ frames were co-added spearately in each of the two photometric bands using the SWarp software \citep{Bertin2003}. 
Individual R and K$_s$ frames were resampled to a pixel scale of 0.29\arcsec/px and rescaled to the same photometric zero-point before co-adding.
A mosaic was created for each photometric band averaging the VIMOS or SOFI frames.
Source detection and photometry were performed in the R-band using SExtractor \citep{Bertin1996}. 
Photometry in the K$_s$ band was obtained for the sources detected in the R band.
Photometric zero-points relative to Vega were calculated for the R-band using the photometric observations reported by \citet{LaFranca2004}.
Photometric zero-points for the J and K$_s$ band are from Dias et al., (in preparation).  
To account for possible K$_s$ sources that do not have an R-band counterpart, detection and photometry were also performed in the K$_s$ band individually
(there are 6 counterparts detected in the K band without detection in the R band).

\subsection{Optical spectroscopy}
The XMM ELAIS-S1 field was the target of several spectroscopic campaigns in 2004-2006 . 
Spectroscopy of 13 optically bright (R$<$20) sources was performed with the EFOSC spectrograph at ESO/3.6m telescope in October 2005, covering the 4000-9000 $\AA$ wavelength range with grisms \#6 and \#13.

Spectroscopic targets with a limiting magnitude of \R=24 were observed with VIMOS/VLT.
These observations were carried out in the  multi-object spectroscopy (MOS) mode, with the Low Resolution Red (LRR) grism ($\lambda/\Delta\lambda \sim 210$),  covering the 5500-9500 \AA~ wavelength range, with 1 to 4 hours exposure time.
During the two VIMOS observing runs, in 2005 and 2006, 196 spectra of X-ray sources have been obtained. 
In addition, about 2000 spectra of 24 \micron, 8 \micron~ and K selected sources were collected. The latter spectra will be reported in a separate publications by Sacchi et al. (2008).
VIMOS data reduction was carried out using the VIMOS Interactive Pipeline and Graphical Interface (VIPGI) \citep{vipgi}.  

Spectroscopy of optically faint X-ray sources, and in particular sources with high X/O, has been carried on with FORS2/VLT in 2006-2007.
Spectra of 45 faint sources (R up to $\sim$25) were collected using the 150I grism
of FORS2/VLT MXU with exposure times of 2.7 hours each, covering the wavelength range 3700-10300 $\AA$.
FORS2/VLT data (together with ESO/3.6m) were reduced using standard IRAF procedures.
Spectra of 9 counterparts are drawn from \citet{LaFranca2004}.
We obtained in total 263 optical spectra of counterparts of X-ray sources.   
Figure \ref{radec}  shows the distribution of the X-ray sources in the sky together with the positions of the \chandra~ pointings. 
The different symbols used in the plot reflect the classification of the sources (see section \ref{spec}).
The density of the spectroscopic targets is not uniform
across the field. In particular the two left quadrants 45\perc~ of the sources are identified, while in the right ones the fraction is $\sim54$\perc.

\begin{figure}
\centering
\begin{tabular}{c}
\includegraphics[width=8.5cm]{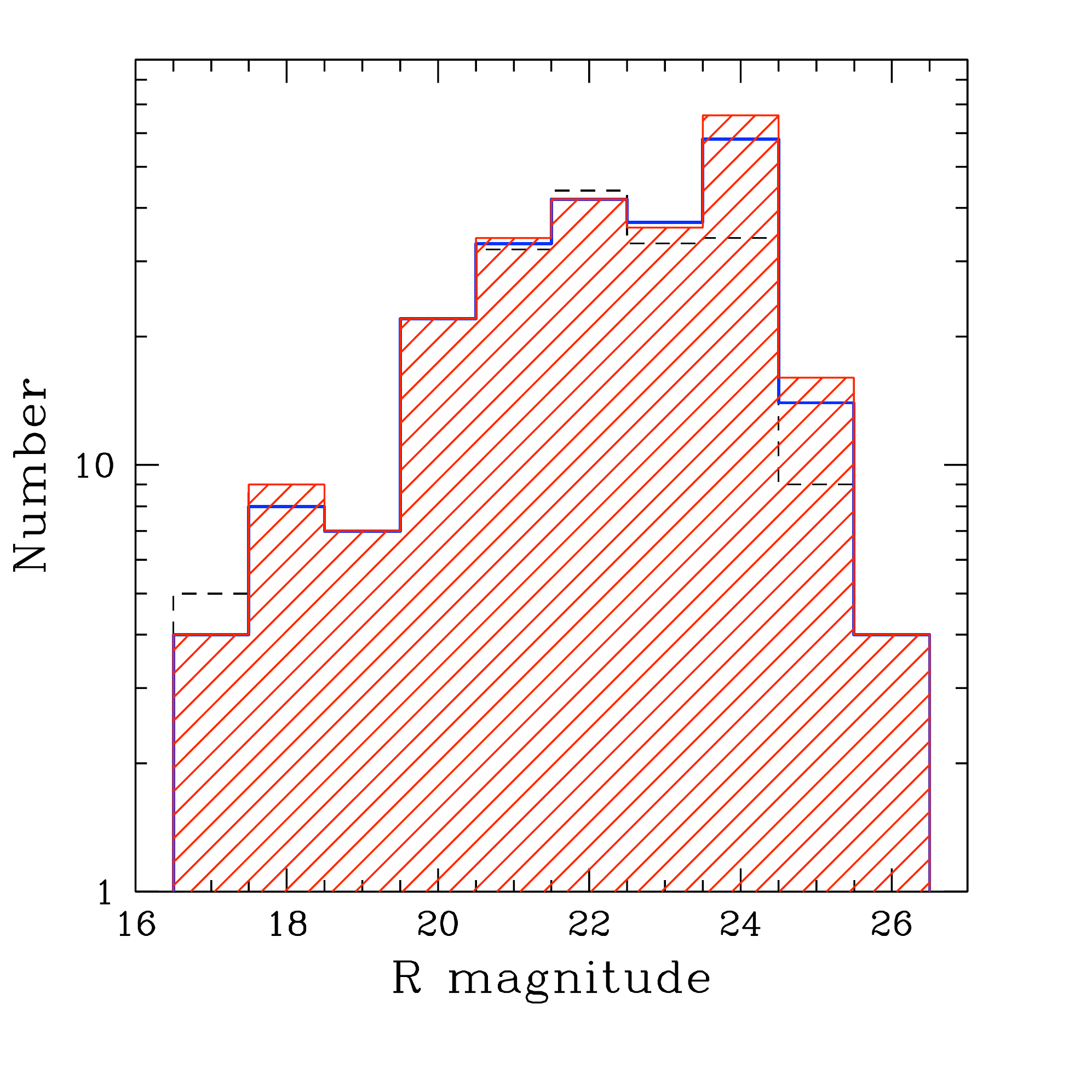}\\
\end{tabular}
\caption{\R~ magnitude distribution of the counterparts detected by \chandra. The identifications obtained using the optical and infrared source counts (black dashed and blue solid histograms, respectively) are compared with the identifications given by the \chandra~ positions (red shaded histogram).
}
\label{istid}
\end{figure}

\section{Source identification}
Since \chandra~ positions are available only for a fraction of XMM sources, we
have first identified the optical/infrared counterparts independently of \chandra~ data.
We have then compared these identifications with the \chandra~ positions for common sources to assess their validity. 
In this way we have identified cases where
\chandra~ observations are crucial to assign the correct counterpart to the X-ray source.

The typical \xmm~ error circles ($\approx 6\arcsec$ radius at 99\% confidence level, \citet{Fiore2003}) are likely to contain several faint optical and infrared sources.
Therefore, we follow a statistical approach to assign the most probable optical/infrared counterpart to each X-ray source.
For the optical band, 
we calculated the chance coincidence probability for each optical source contained inside the 6\arcsec~ error-circle  (see e.g. \citet{Brusa2007}), using the distance between the \xmm~  and the optical positions ($d_{X-O}$), and the R-band galaxy counts N(R): 

$$ P(i) = exp[- \pi d_{X-O}^2  N(R)] $$
  
where $N(R)$ are the integral source counts evaluated directly from the VIMOS photometry.
The best fit relation is $log N \propto 0.267 \times R$.

For the 3.6 \micron~ band we applied the same equation, but using the 3.6 $\mu m$ source counts from \citet{Fazio2004}, and the distance between the \xmm~  and the infrared positions. 
The identification based on IRAC 3.6 \micron~ provides a counterpart different from the most probable optical one in 67 over 478 cases (15\perc).
All of them are optically faint sources: 67\perc~ have mean R$\sim$23.3,  and the remaining do not have an optical detection down to R$\sim$24.5.
  
For 31 error-boxes (6\perc~) we find multiple optical and infrared counterparts with similar probabilities.
In these cases we identified the X-ray source with the counterpart with the brightest MIPS 24 \micron~ flux
(183 sources ($\sim$38\perc) are detected by MIPS at 24 \micron~ down to 400 $\mu$Jy). 
In addition, we visually inspected the sources one by one and used 
the spectroscopic information where available to validate the identification.

Finally, we assigned to each X-ray source either an optical or infrared counterpart.
Most of them (92\perc) are detected in the R band, while the remaining are detected in IRAC. 
420 X-ray sources (88\perc~ of the sample) have a counterpart detected both in the optical and in IRAC.

\begin{figure}[t]
\centering
\begin{tabular}{c}
\includegraphics[width=8.5cm]{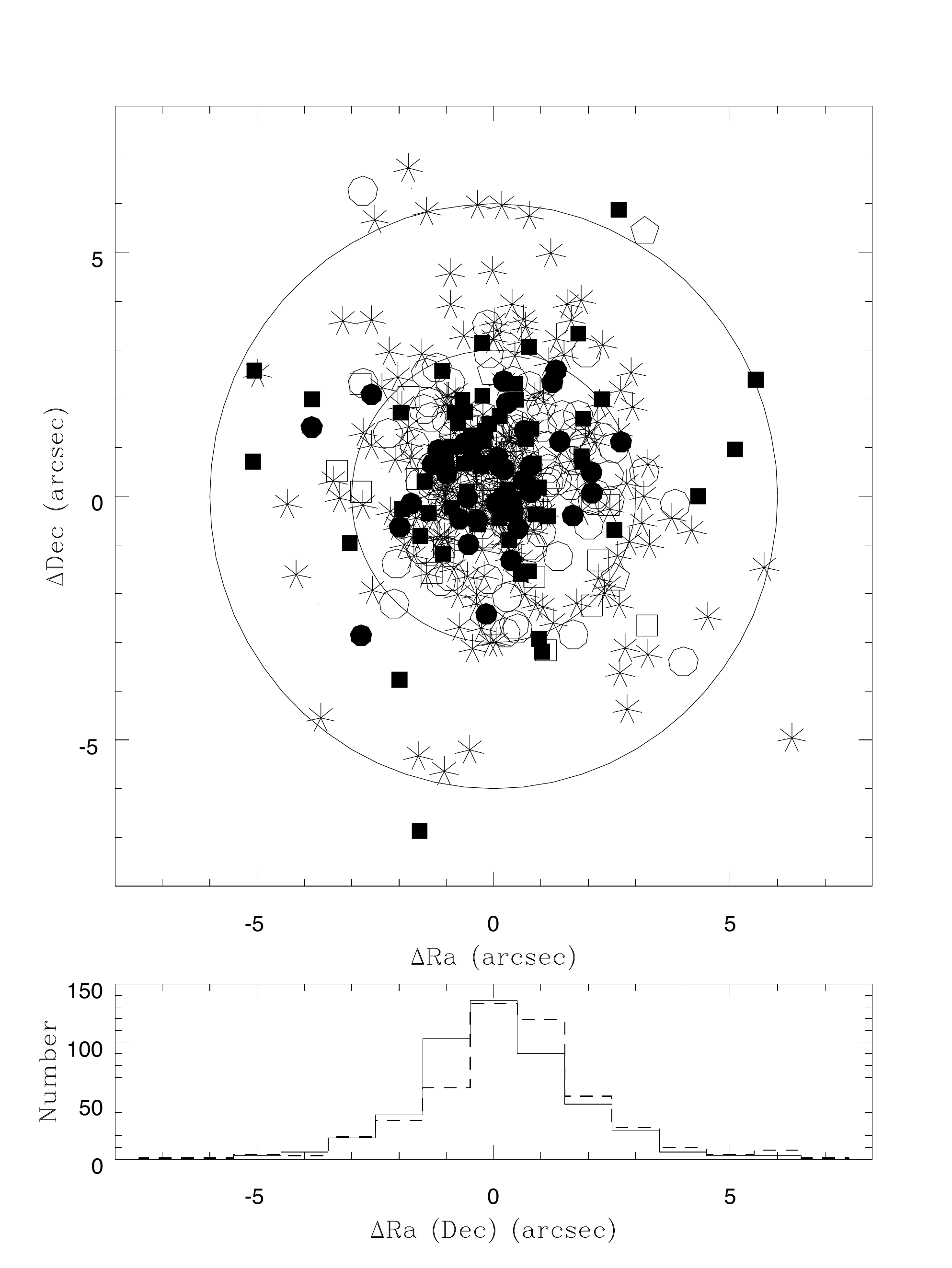}\\
\end{tabular}
\caption{The offsets between the XMM and the optical/infrared counterpart positions. Same symbols as in figure \ref{radec}. The circles represent the 3\arcsec~ and 6\arcsec~ \xmm~ error-boxes. Lower panel: offset histograms for $\Delta$Ra (solid histogram) and $\Delta$Dec (dashed histogram).  }
\label{radec1}
\end{figure}

\subsection{Chandra positions}
About 78\perc~ of the \chandra~ positions are within 3\arcsec~ from the \xmm~ centroid.
The \chandra~ detections confirm the  counterpart identified through the 3.6 \micron~ chance coincidence probability for 96\perc~ of the sample, while yields 4\perc~ of new identifications (10 sources).
In half of the cases the IRAC sources are strongly blended. In the other cases \chandra~ chooses one of two sources with similar probability.
Nine of the ten sources have faint optical counterparts (R$>$24). 
In one case it coincides with two bright stars closer than a few arcsecs.

As an exercise, we compared the \chandra~ detections with the identifications provided by optical photometry only. 
\chandra~ confirms the optical identification in 197 cases, while it identifies a new counterpart for 50 sources (20\% of the sample).
Taking into account only the optically faint sources (R$>$23), the rate of new identifications increases to $\sim$40\% (44 new identifications over 109 sources).

The  \R~ magnitude distribution of the counterparts detected by \chandra~  is shown in Figure \ref{istid}.

Figure \ref{radec1} shows the displacement between the \xmm~ centroid and the optical/infrared position of the identified counterparts. 
The different symbols used in the plot reflect the classification of the sources (see section \ref{spec}).
98\perc~ of the counterparts are located within 6\arcsec~ from the XMM position.
The median (and inter-quartile range shown in brackets from now on) of the displacement $d_{X-O}$ are 1.74\arcsec [0.86\arcsec],
with 78\perc~ of the optical/infrared counterparts (374 sources) falling within 3\arcsec~ from the X-ray position, consistent with
other surveys (HELLAS2XMM, \citet{Fiore2003}, COSMOS, \citet{Brusa2007}). 
The sources with $d_{X-O}>$6\arcsec~ (11 in total) are nearby galaxies extended in the optical. 
There is one broad-line AGN with $d_{X-O}>$6\arcsec, for which the XMM position may be due to the mixed contribution of a nearby extended galaxy and to the AGN itself.
On average, broad-line AGNs are more concentrated than the full sample on the X-ray positions, 
showing a median displacement $d_{X-O}=$1.42\arcsec [0.66\arcsec], and 93\perc~ of them are located within 3\arcsec.
The median displacement of the 93 faint (R$>$23), spectroscopically unidentified sources is 2.2\arcsec [1.0\arcsec]. 
Among these sources, those with neither \chandra~ nor \spitzer~ detection, and 
$d_{X-O}>$3\arcsec (6 in total) may be false identifications.

\begin{figure*}
\centering
\begin{tabular}{cc}
\includegraphics[width=8.5cm]{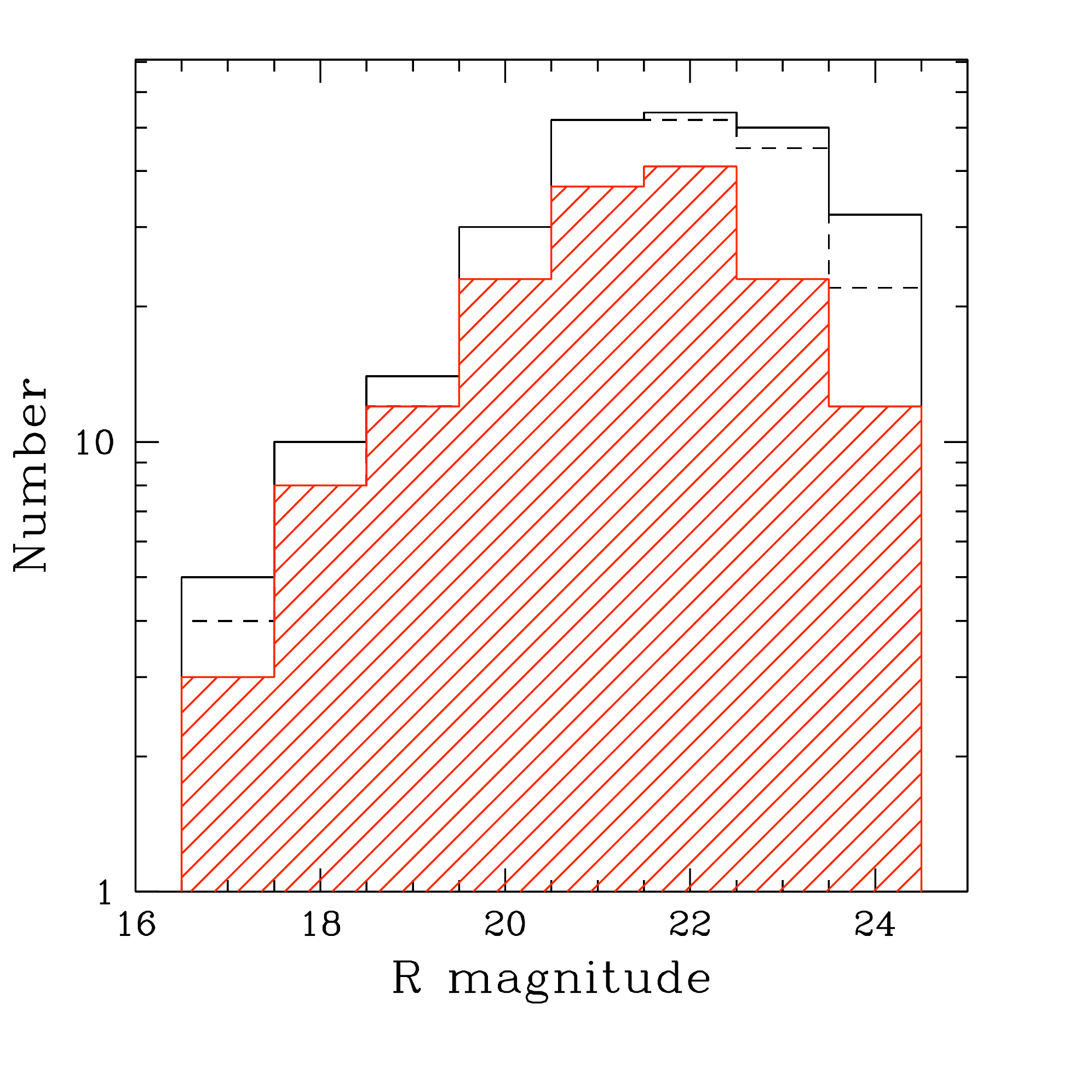} 
\includegraphics[width=8.5cm]{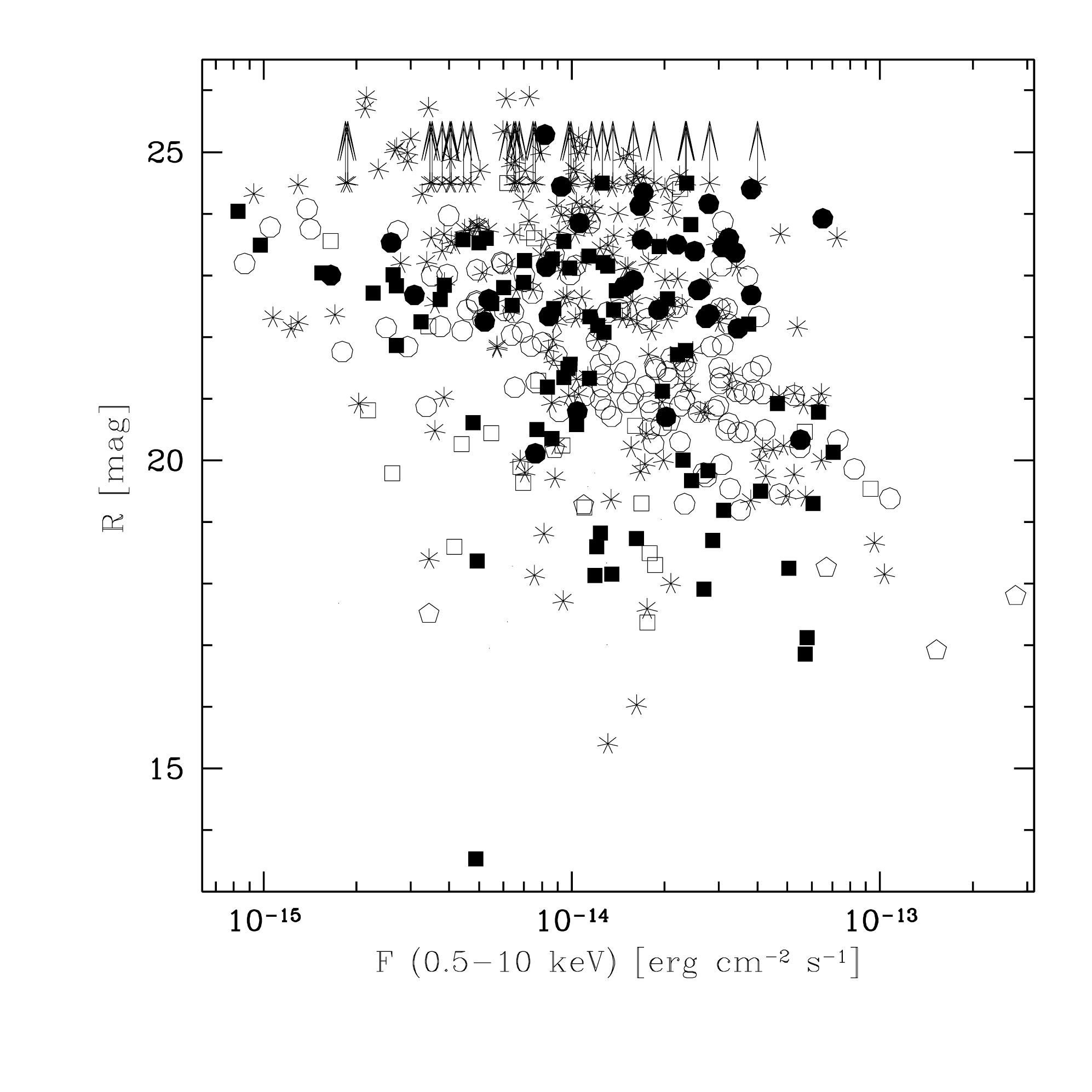}
\end{tabular}
\caption{Left panel: redshift quality distribution in \R~ magnitude  bins. The solid histogram represents the whole spectroscopic sample, the dashed and shaded histogram the safe redshifts (quality 1 and 2 respectively).
Right panel: R-band magnitude vs. 0.5-10 keV flux. Symbols as in figure \ref{radec}. }
\label{qualz}
\end{figure*}

\section{Results}

\subsection{Spectroscopic redshifts and source classification}\label{spec}
We collected 263 spectra of counterparts of X-ray  sources.
Redshifts were computed using the \emph{rvidlines} IRAF tool. The average of the redshifts corresponding to the observed line centers gives the source redshift.
Depending on the number of spectral features used to derive it, we assigned a quality flag to each redshift. 
Objects identified through at least two emission (absorption) lines were assigned quality 2. 
Spectra that showed one clearly recognizable feature were assigned quality 1.  
We were able to assign a secure redshift (quality$>=$1) to 237 sources (90\perc~ of the spectroscopic sample).
Less secure determinations (i.e. quality $<1$) are found in 10\perc~ of the cases, and 60\perc~ of them has an optical counterpart fainter than R$=$23 and no prominent emission lines in its spectrum. 
Figure \ref{qualz} shows the redshift quality in different \R~ magnitude  bins. 
The fraction of secure redshift is 94\perc~ for \R$<23$, while drops to 76\perc~ for sources with \R$>$23. 
In the following analysis we use redshifts with quality $\ge$1 (237 sources).
Figure \ref{qualz} (right panel) shows the R-band magnitude vs. the 0.5-10 keV flux.
The spectroscopic completeness is $\sim 50$\perc, while it is 65\perc~ for $R\le23.0$.

We classified the sources according to their optical spectral features in five broad classes:
 
 \begin{itemize}
 \item Broad-line AGN (BL AGN): sources with broad (FWHM $>$ 2000 km s$^{-1}$) emission lines such as \ion{C}{iv} $\lambda 1549$, \ion{C}{iii]} $\lambda 1909$, \ion{Mg}{ii} $\lambda 2798$, H$\beta$, H$\alpha$.
 
  \item Narrow-line AGN (NL AGN): sources with narrow (FWHM $< 2000$ km s$^{-1}$) high-ionization emission lines indicating the presence of an AGN (\ion{C}{iv} $\lambda 1549$, \ion{C}{iii]} $\lambda 1909$,    \ion{[Ne}{v]} $\lambda 3426$). With the appropriate wavelength coverage, the presence of these emission lines allows classifying sources as AGN2 if broad lines such as \ion{Mg}{ii} $\lambda 2798$ are not detected. 
However in a few cases the small spectral coverage  does not allow classifying as AGN1 or AGN2 sources showing \ion{[Ne}{v]}  emission line.
 
 \item Emission-line galaxies (ELG): sources with narrow emission lines, but no AGN signature in the optical spectra.
They show strong low-ionization emission lines, such as \ion{O}{ii}  $\lambda 3727$, H$\beta$, \ion{O}{iii} $\lambda \lambda 5007,4959$, H$\alpha$, that may be produced by thermal photons from hot stars. They often show also CaHK $\lambda \lambda 3933,3969$ absorption and the continuum break at 4000 $\AA$. 
This class is likely to include a fraction of NL AGNs, misclassified due to the 
small wavelength range or to low S/N spectra.
  
  \item Absorption-line galaxies (GAL): sources with spectra typical of early-  galaxies, characterized  by absorption features such as CaHK $\lambda \lambda 3933,3969$ and the 4000 $\AA$ continuum break.
  
 \item Stars: these may be either X-ray emitting stars or chance coincidences, the X-ray emission arising from an undetected background source. These sources will be not be discussed in the following.
 \end{itemize}

\begin{figure}
\centering
\begin{tabular}{c}
\includegraphics[width=8.5cm]{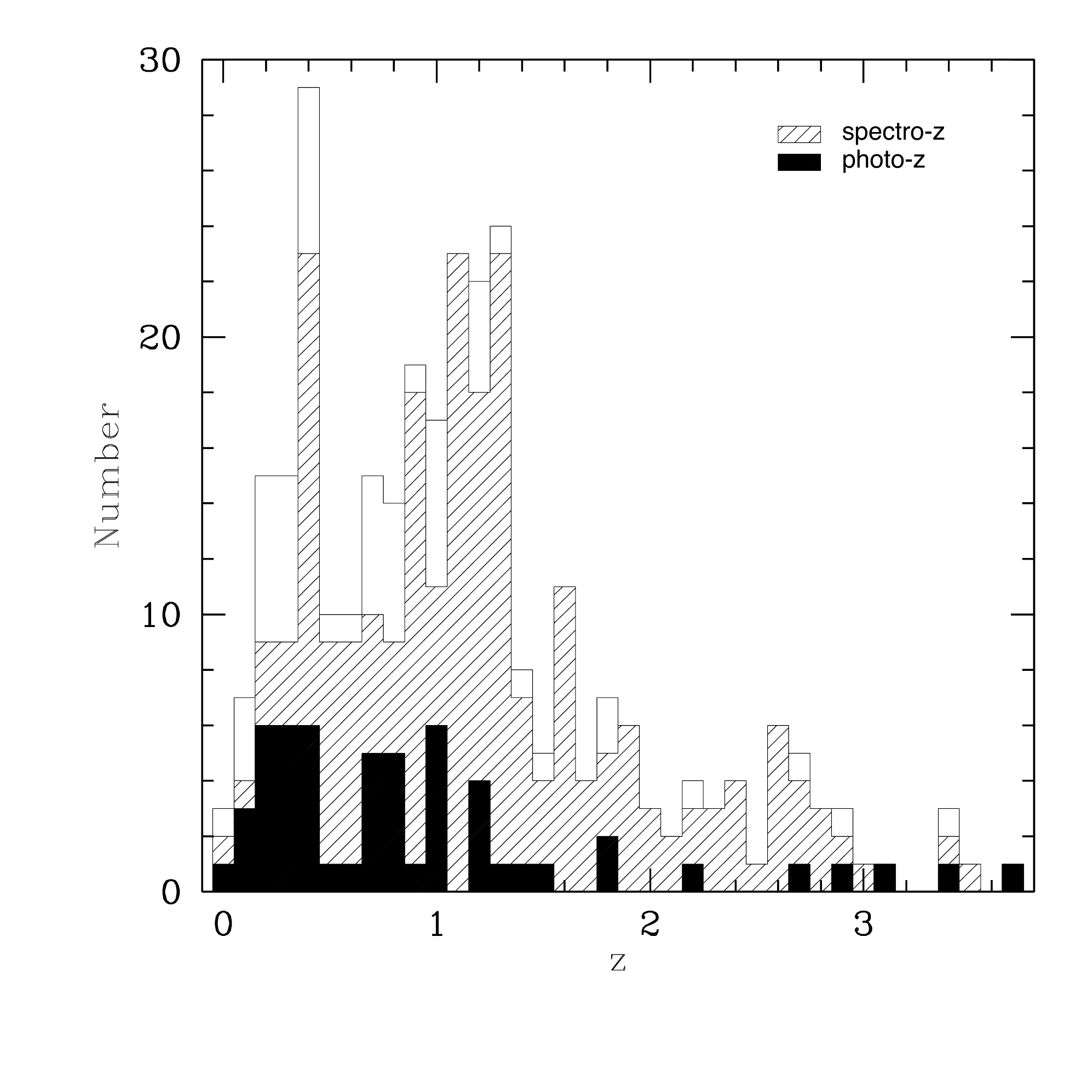}\\
\includegraphics[width=8.5cm]{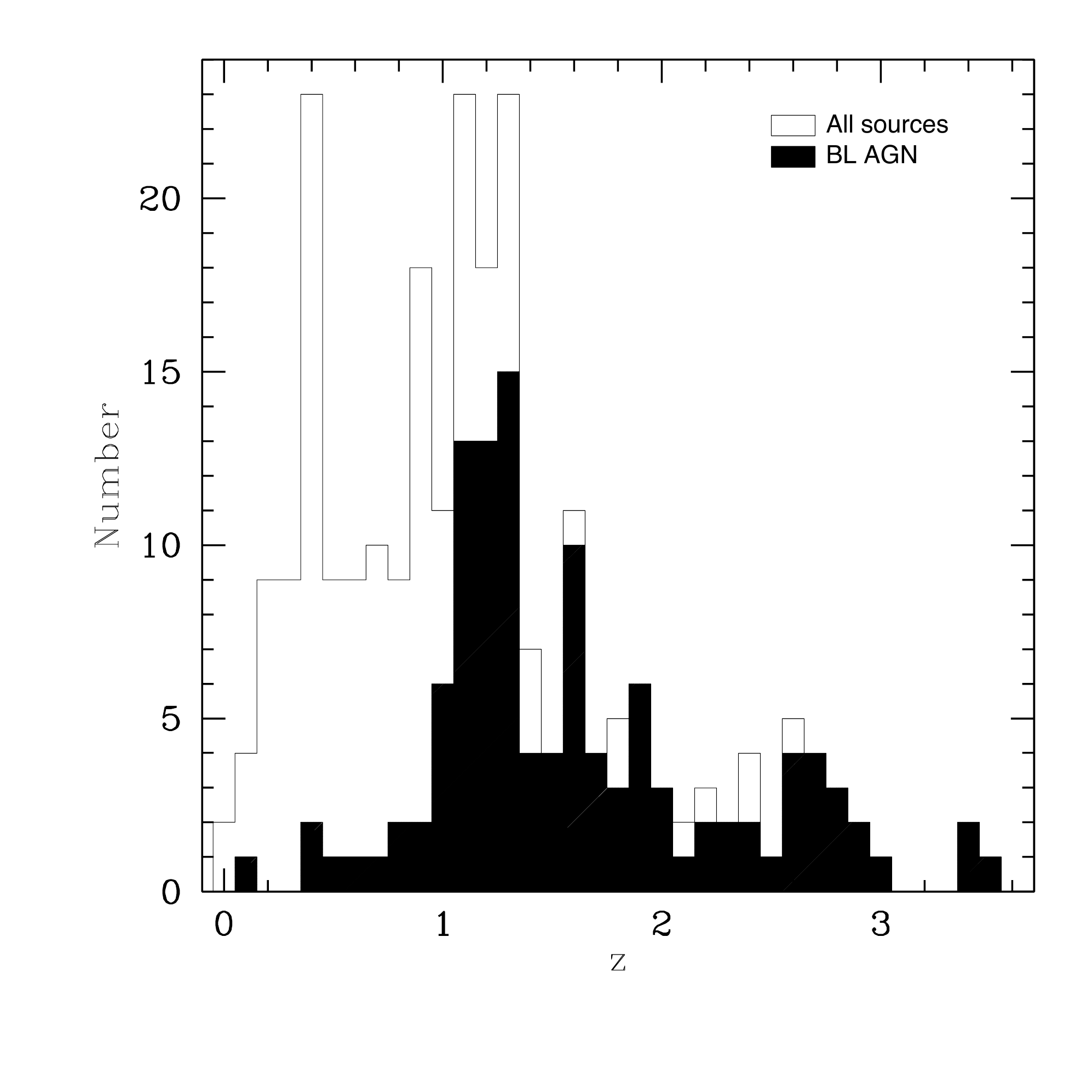}\\
\end{tabular}
\caption{Top panel: redshift distribution of the X-ray sample (open histogram). Spectroscopic and photometric redshifts (see section \ref{photoz}) are represented by the shaded and the filled histograms, respectively.
Bottom panel: redshift distribution for the spectroscopic sample with z$>$0 (open histogram). The black filled histogram shows the  contribution of broad-line AGNs.}
\label{istz}
\end{figure}

\begin{figure*}
\centering
\begin{tabular}{cc}
\includegraphics[width=8.2cm]{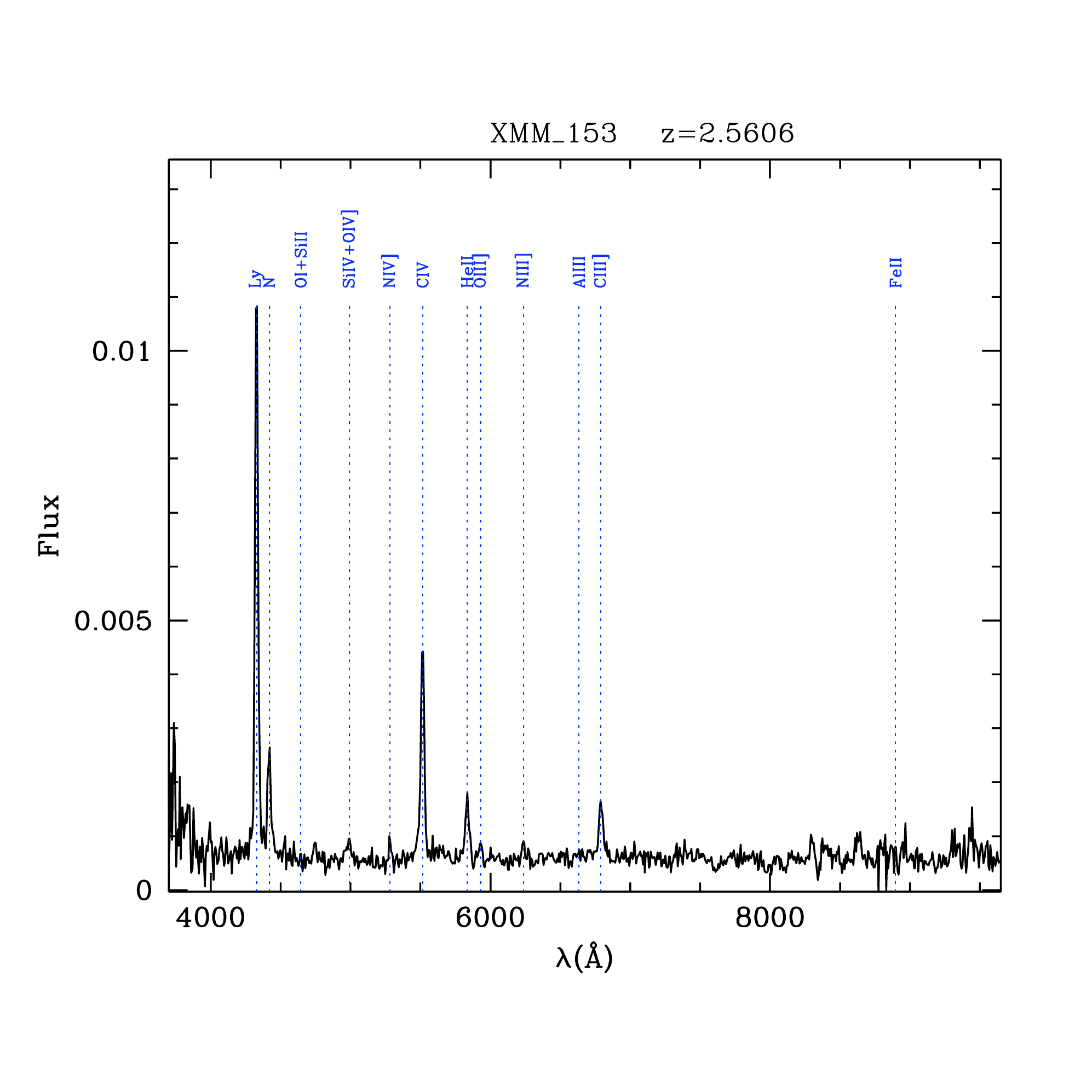}
\includegraphics[width=8.2cm]{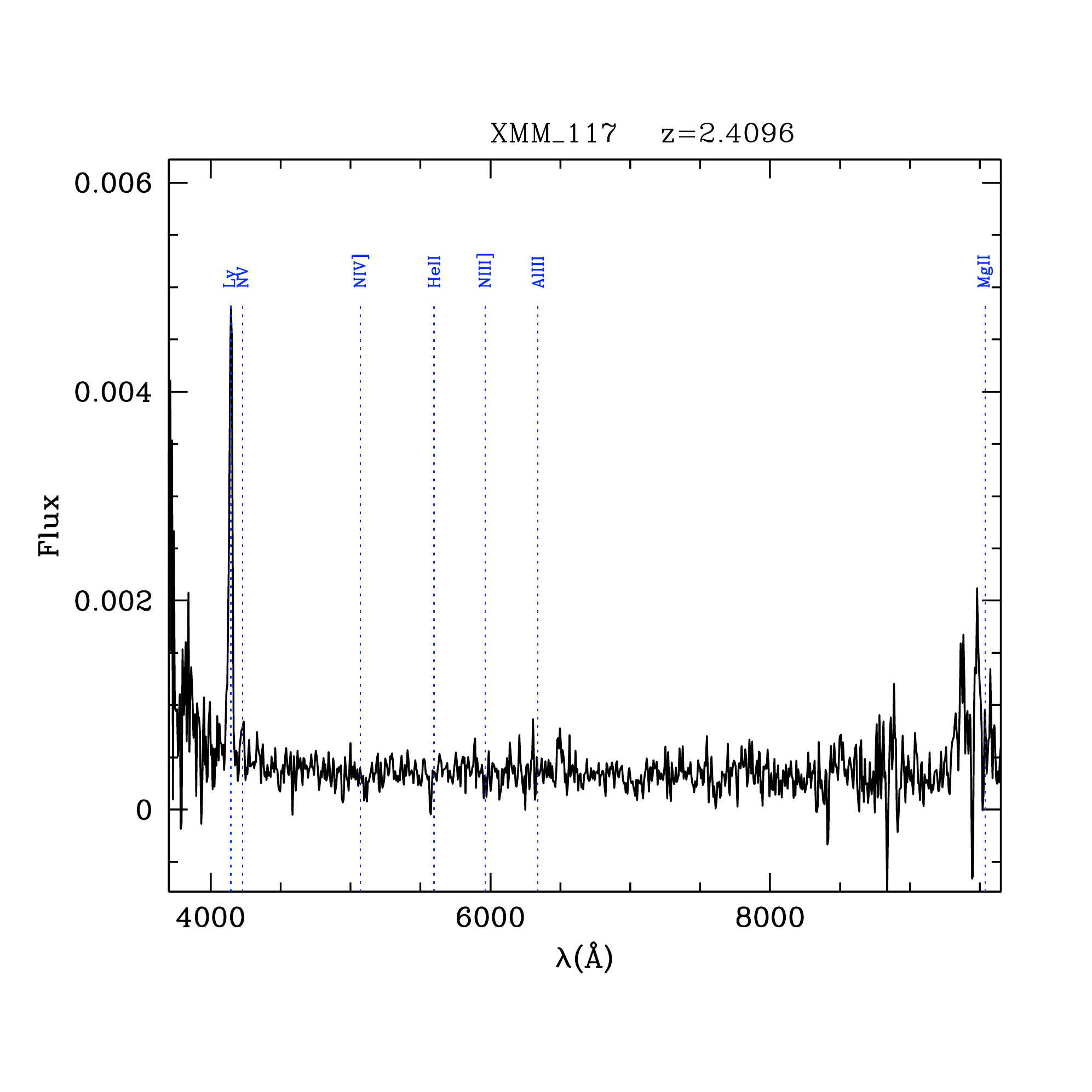}\\
\includegraphics[width=8.2cm]{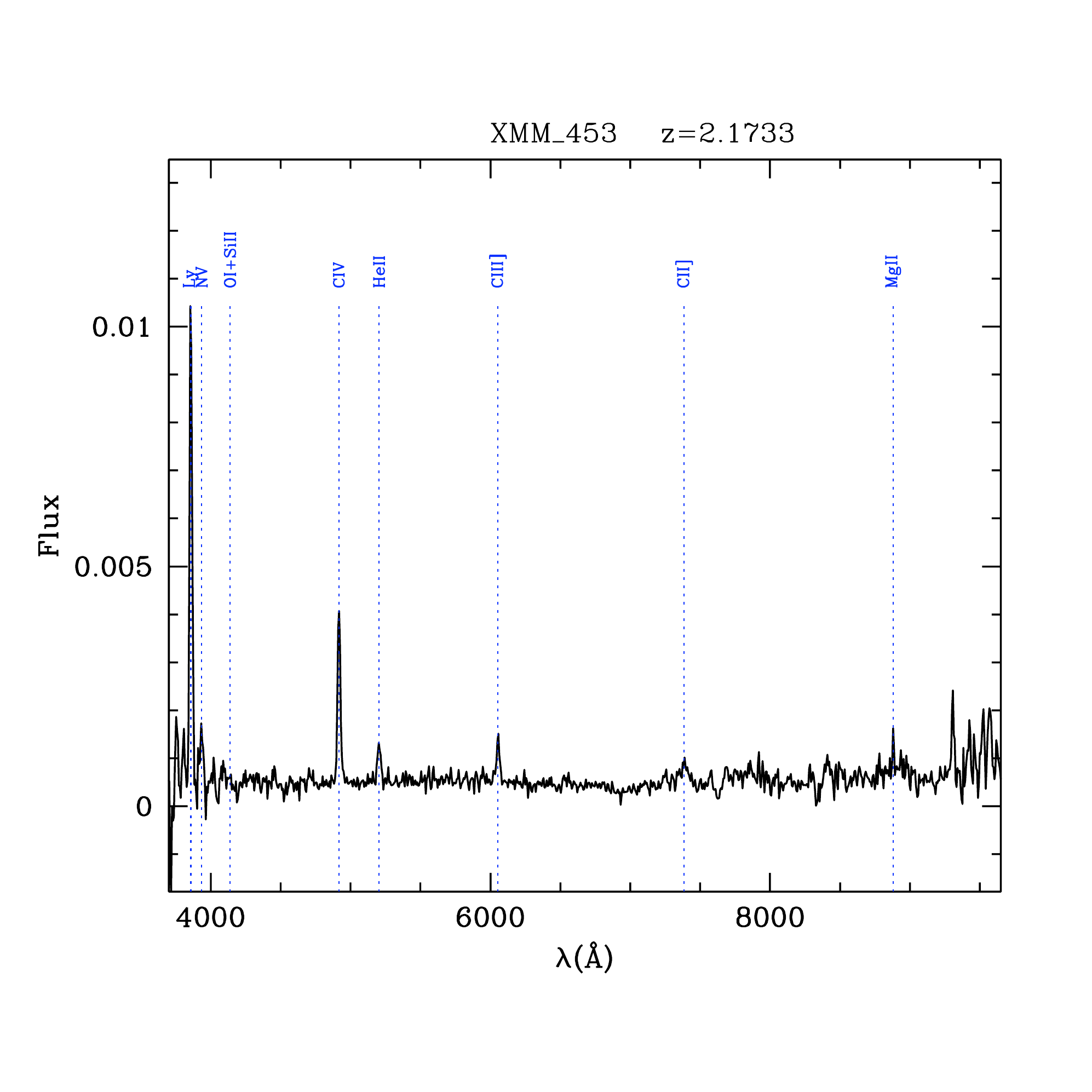}
\includegraphics[width=8.2cm]{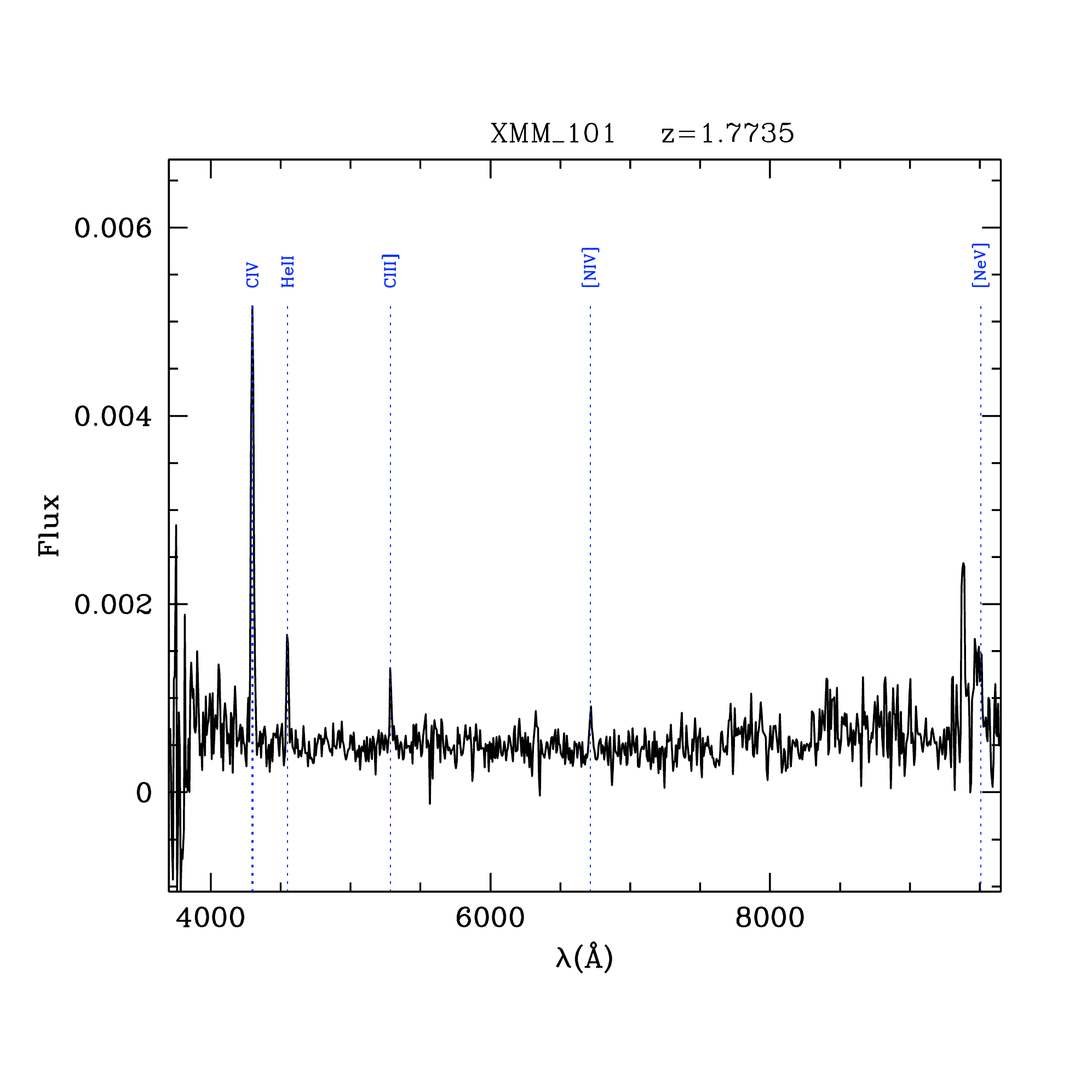}\\
\includegraphics[width=8.2cm]{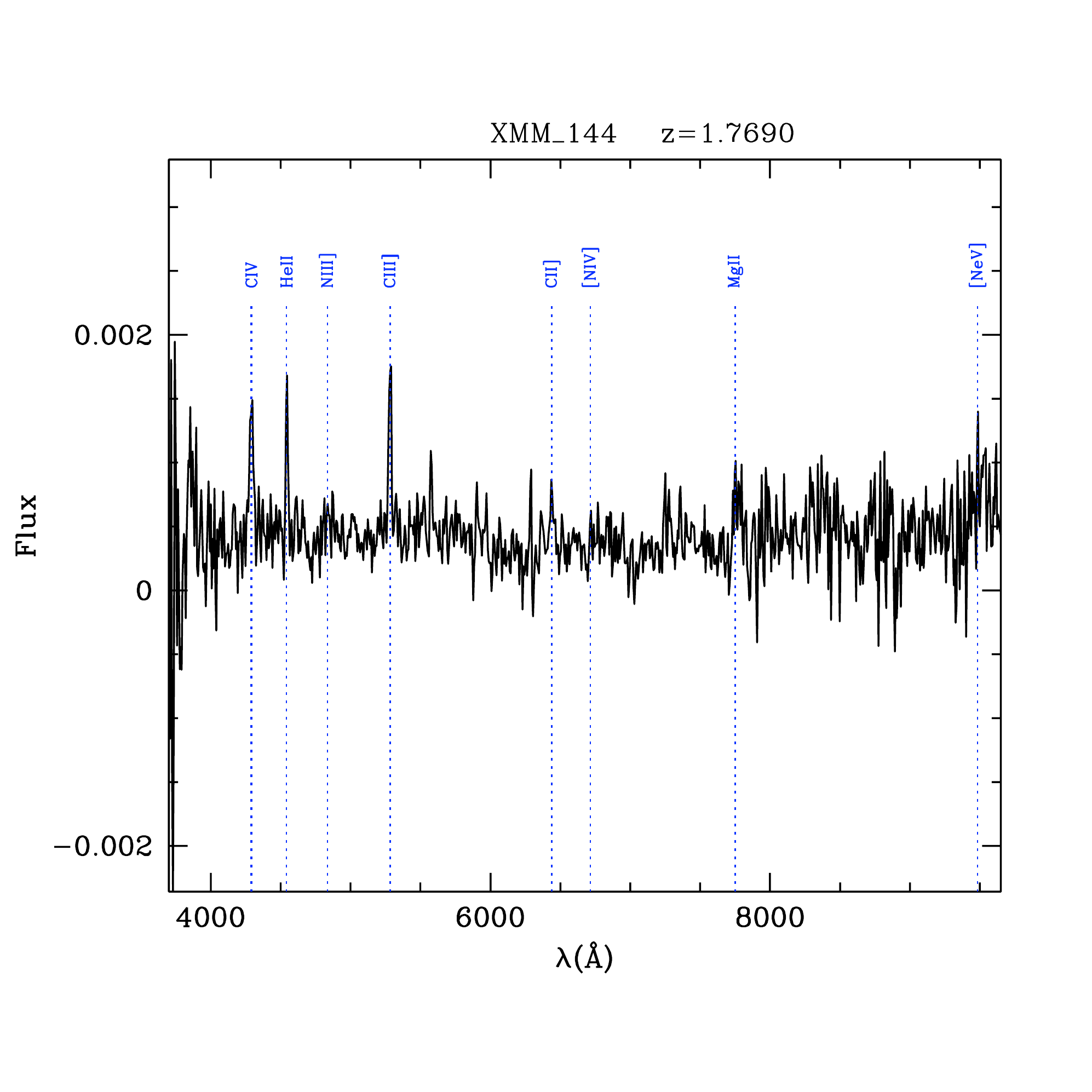}
\includegraphics[width=8.2cm]{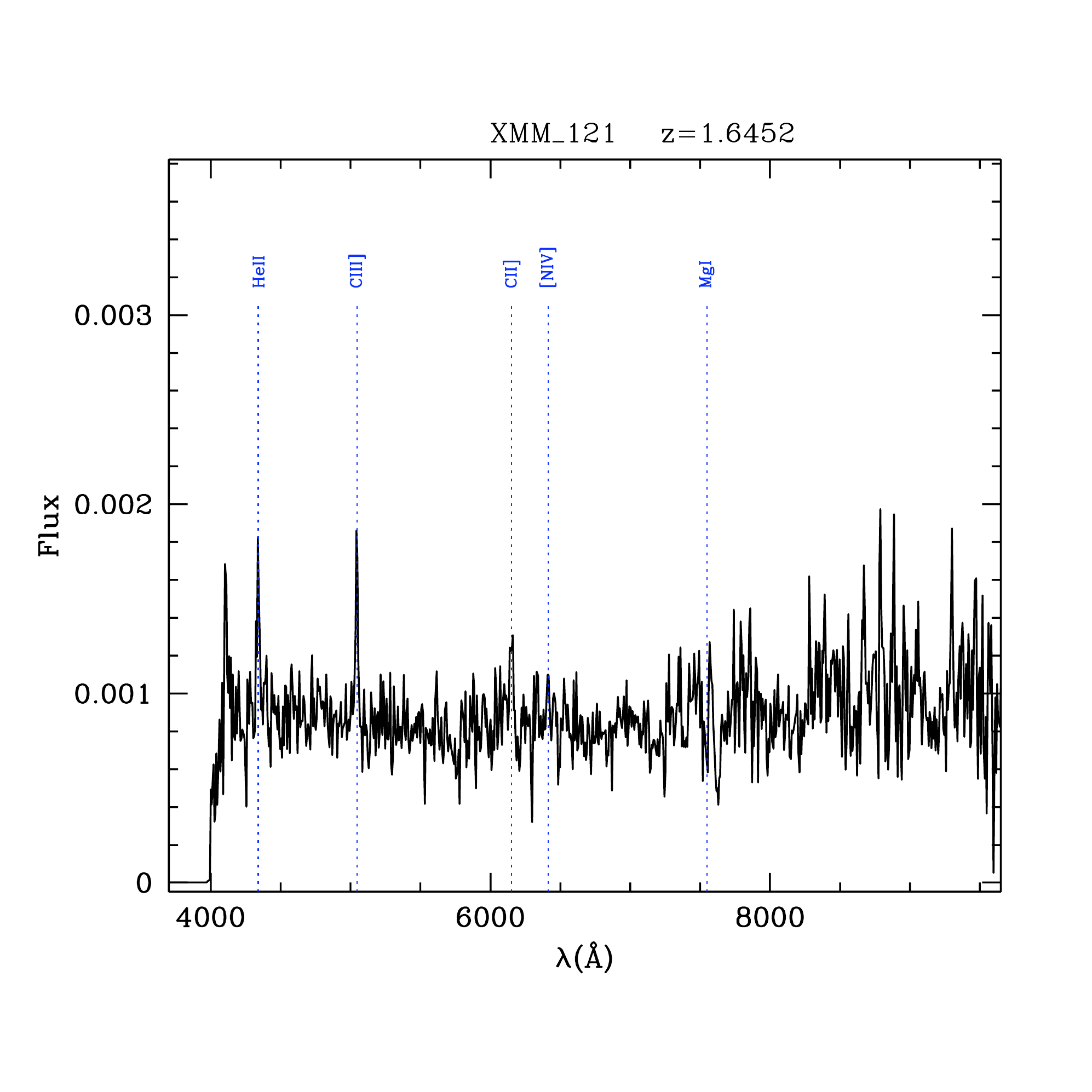}\\
\end{tabular}
\caption{Optical spectra of six narrow-line AGNs with z$>1.6$. The source redshift and identified  spectral features are also marked.}
\label{spettri}
\end{figure*}

\begin{figure}
\centering
\begin{tabular}{c}
\includegraphics[width=8.5cm]{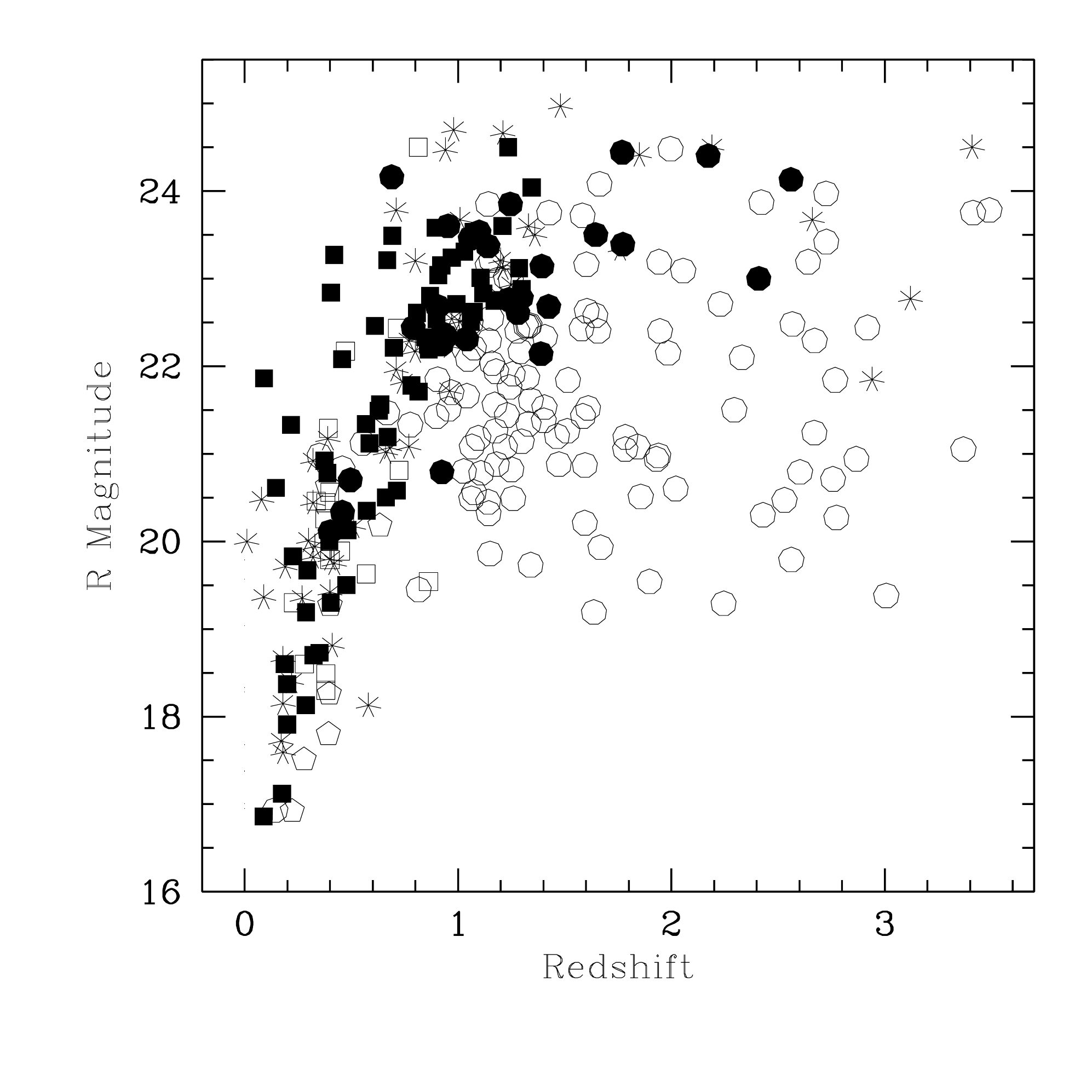}\\
\includegraphics[width=8.5cm]{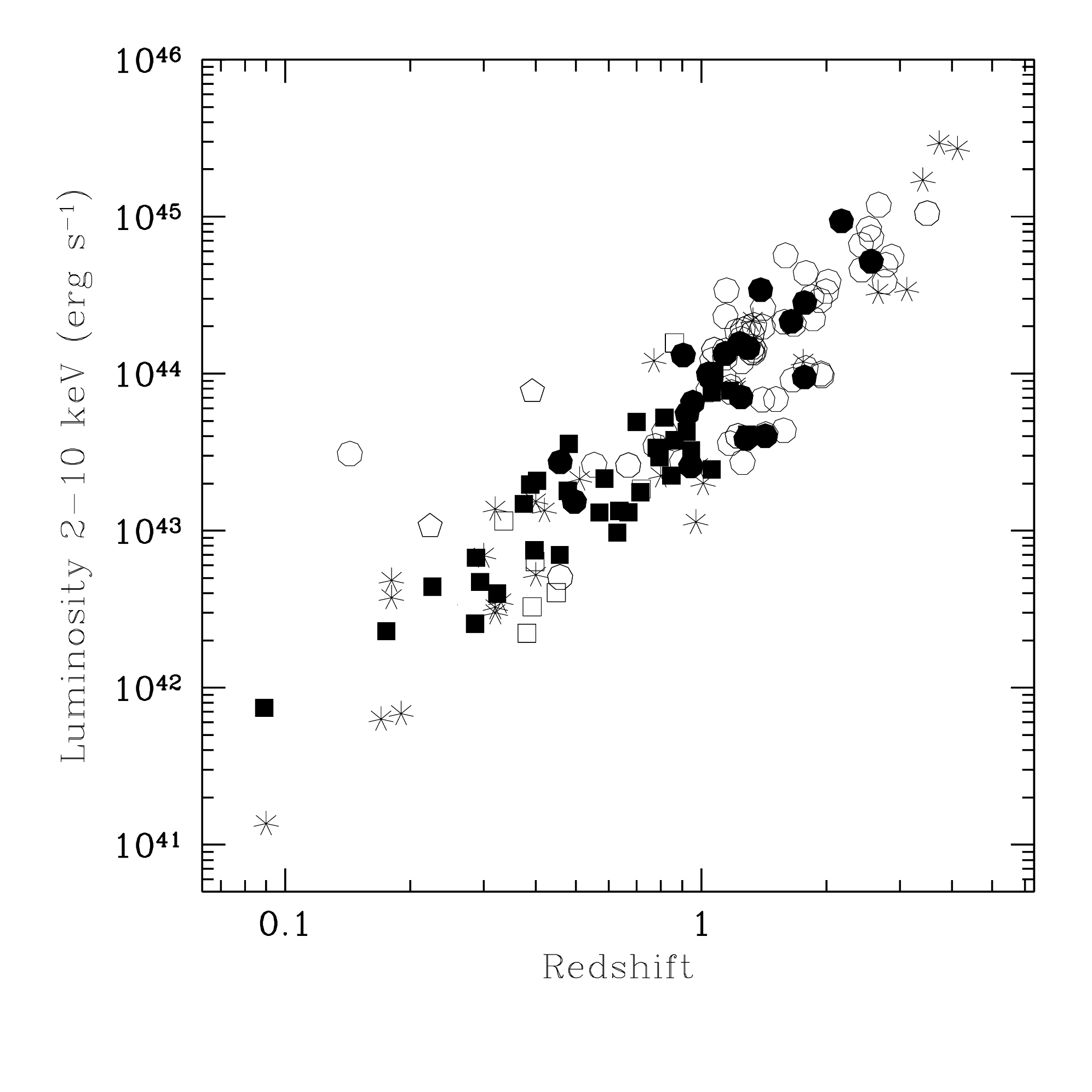}
\end{tabular}\caption{Upper panel: \R~ magnitude versus redshift distribution for the spectroscopic sample. Same symbols as in Figure \ref{radec1}. Lower panel: 2-10 keV luminosity versus redshift. 
}
\label{zr}
\end{figure}

Among the spectroscopically classified sources we find: 111 BL AGNs (47\perc~ of the sample with spectroscopic redshifts),  28 narrow-line AGNs (12\perc), 63 emission-line galaxies (26\perc)
and 18 absorption-line galaxies (8\perc).
NOT BL AGNs, i.e.  all sources classified as AGN based on their X-ray luminosity, unresolved X-ray emission and z$>$0, without visible broad lines in their optical spectra,  make up 46\perc~ of the sample.
The 6 galaxies corresponding to the X-ray extended sources, not included in the previous breakdown, will be discussed separately in section \ref{ext_sec}.
The remaining 11 sources (5\perc~ of the sample) are stars.

The redshift distribution for the X-ray sample is shown in Figure \ref{istz}. 
The distributions of the spectroscopic (shaded) and photometric redshifts (black filled histogram, see section \ref{photoz}) are shown in the top panel.
Figure \ref{istz}, lower panel, shows the redshift distribution of the extragalactic spectroscopic sample (open histogram), and the contribution of the BL AGNs (filled histogram).  
The BL AGN population shows a broad redshift distribution, with median redshift $z_{BLAGN}$= 1.40 [0.39]. 
The observed fraction of BL AGN increases with redshift, reaching 90\% of the whole spectroscopic population at z$>1.5$.
This class of objects have relatively bright optical counterparts ( median R= 21.5 [0.8]) and prominent emission lines in their spectra, making them easy to identify over a large redshift range.
We find 28 narrow-line AGNs, 40\perc~ of which are type 2 QSOs, having log\Lx$>43.8$.
They are found up to z$\sim$2.6, and are associated with fainter optical counterparts (median R=22.8 [0.6]). 
Figure \ref{spettri} shows a sample of type 2 QSO spectra.
The population of emission-line galaxies dominates the identifications at z$<$1, with a median z=0.67[0.27]. 
The sample of sources classified as ELG may include a fraction of NL AGNs, which may have been misclassified due to the limited wavelength range. 
The decline of narrow-line sources  and normal galaxies above $z\sim1$ is likely to be due to selection effects \citep{Brusa2007}. 
In particular, NL AGNs are difficult to identify spectroscopically at high redshift, being on average fainter in the R band and therefore beyond the spectroscopic limit of our sample (\R$\sim24$).
This selection effect is shown in Figure \ref{zr}, which presents a plot of \R~ magnitude versus redshift. 
The rapid rise of \R~ magnitude with redshift for NOT BL sources (filled symbols) is due to their relatively narrow range in absolute magnitude.
Conversely, BL AGNs show a broad \R~ magnitude distribution over a large redshift range, and dominate the identifications at high luminosity (\Lx $\ge 10^{44}$ \ergs, Figure \ref{zr} bottom panel).
The absorbed X-ray luminosities, \Lx, were derived assuming a power-law spectrum (F($\nu) \propto \nu^{-\alpha_x}$) with spectral index $\alpha_x$=0.8.
In addition, the absence of strong emission lines in the typical wavelength range covered by optical spectroscopy (redshift desert) prevents us from finding ELGs above $z \sim 1.4$ \citep{Eckart2006}.
The fraction of optically obscured (NOT BL) AGNs with log\Lx$\ge 42.0$ \ergs is $50 \pm 6\%$ of all the spectroscopic population, and it is $80\pm 12\%$ in the redshift range z=[0,1.1].
The larger fraction of NOT BL AGN compared to other surveys (e.g. $27\pm7\%$ of HELLAS2XMM, \citet{Cocchia2007}) can be partly due to the faint optical magnitudes reached by FORS2 spectroscopy.
For example, at faint magnitudes (R$\ge 23.2$) we have have spectroscopically identified 31 sources, 2/3 of them being NOT BL AGNs, vs. 10 sources in HELLAS2XMM.

Among the BL AGN detected in the 2-10 keV band, there are 11 sources (19$\pm 6\%$) with an absorbing column $N_H > 10^{22}~cm^{-2}$. 
At a face value this is a larger, but not inconsistent, fraction compared to other surveys \citep{Perola2004}, 
since our $N_H$ values may be overestimated, being derived from the hardness ratio (HR=(Hard-Soft)/(Hard+Soft)).

\begin{figure}[!t]
\centering
\begin{tabular}{c}
\includegraphics[width=8.5cm]{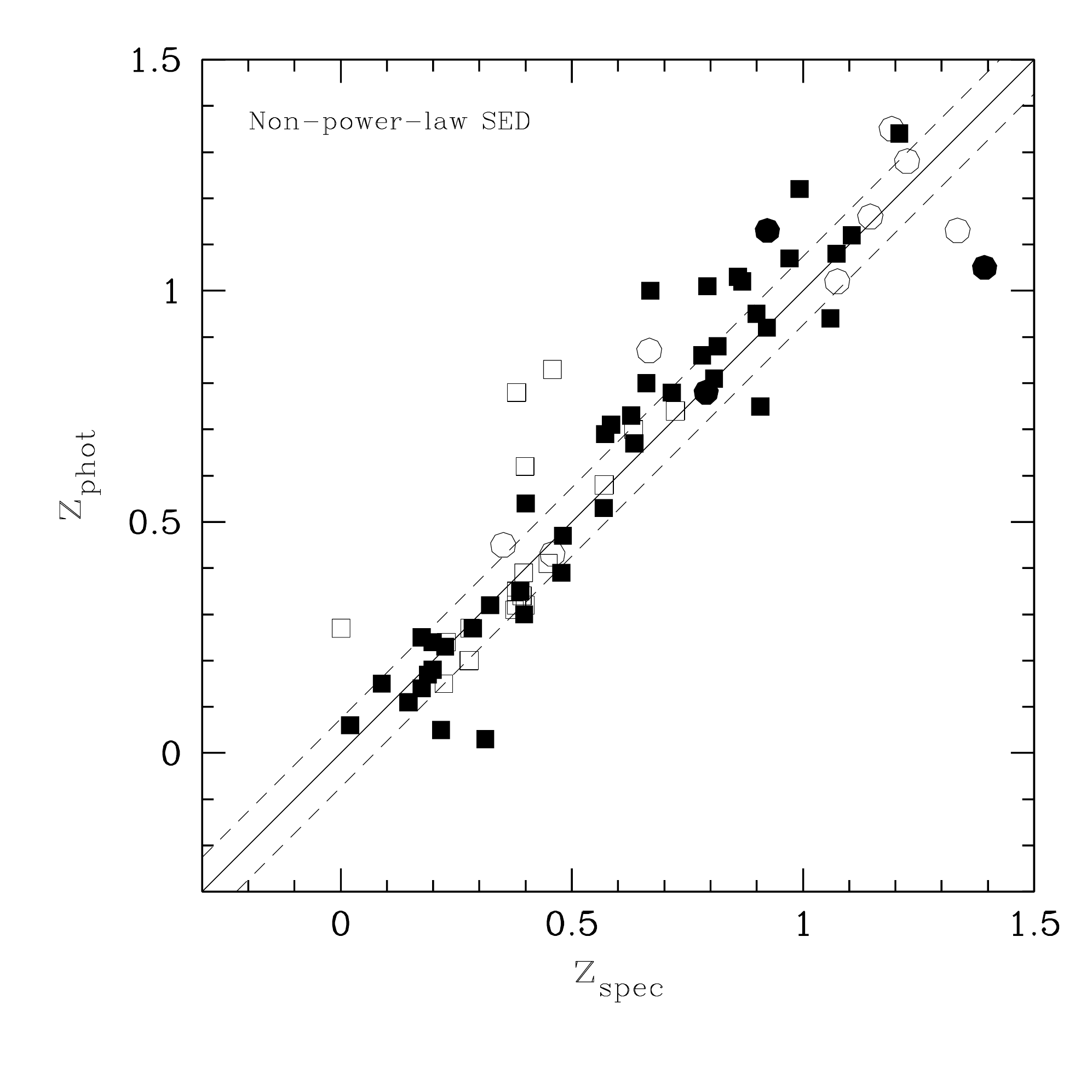}
\end{tabular}
\caption{Photometric versus spectroscopic redshifts for the spectroscopic XMM sources showing a NOT power-law SED. Same symbols as in Figure \ref{radec1}. The solid line is the one-to-one relation, the dashed lines are the 1$\sigma$ deviation.}
\label{photoz_sed}
\end{figure}

\subsection{Photometric redshifts}\label{photoz}
About 53\% of the \xmm~ sources does not have a spectroscopic redshift.
We used the code described in \citet{Fontana2000} and \citet{Fontana2001} 
to obtain photometric redshifts for the full sample of X-ray sources.
Photometric redshifts are calculated by using a $\chi^2$  minimization technique on spectral libraries which include both starburst, passive galaxy  and AGN semi-empirical template spectra from \citet{Polletta2007}, \citet{Fiore2008} and \citet{Pozzi2007}.
Sources with power-law SEDs provided degenerate results (see section \ref{sed} for detailed discussion of SEDs).
Conversely, for the 76 NOT power-law sources with a spectroscopic redshift, the photometric redshift is in fair agreement with the spectroscopic one, producing an average  $\sigma[\Delta z /(1+z)]= $0.087.
The results are shown in Figure \ref{photoz_sed}. 
Therefore, for 68 sources with  a SED dominated by the galaxy
emission, we can consider the photometric redshift as reliable and its value is reported in the catalog.  
Their photo-z distribution is shown in Figure \ref{istz}.
These 68 sources will be analyzed in the following together with the 237 sources with a secure spectroscopic redshift.
For 62 sources without spectroscopic redshift and showing a power-law SED, and for 86 sources with only a few bands above the detection limit,
we were not able to determine a reliable photometric redshift.

\begin{figure}[t]
\centering
\begin{tabular}{cc}
\includegraphics[width=4.2cm]{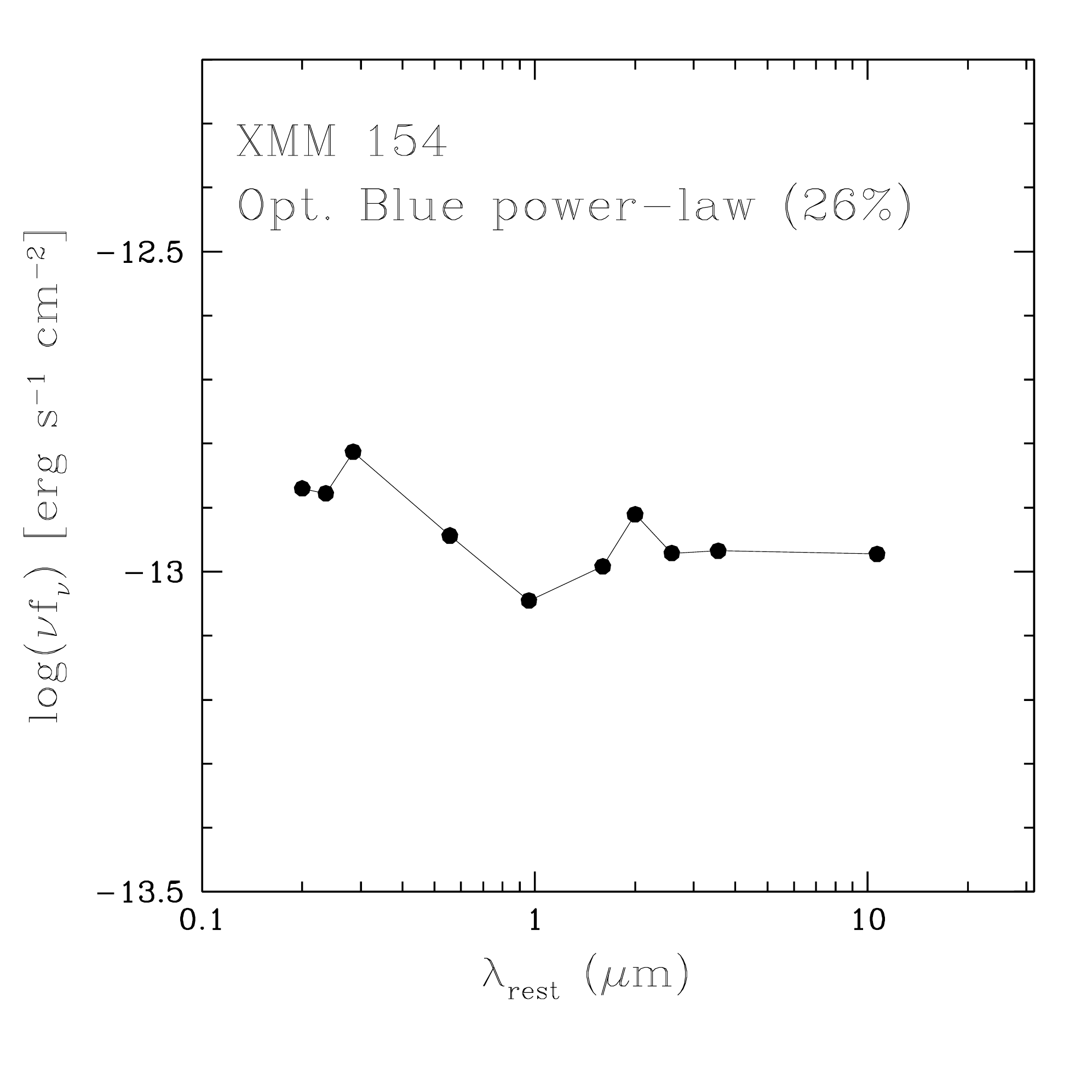}
\includegraphics[width=4.2cm]{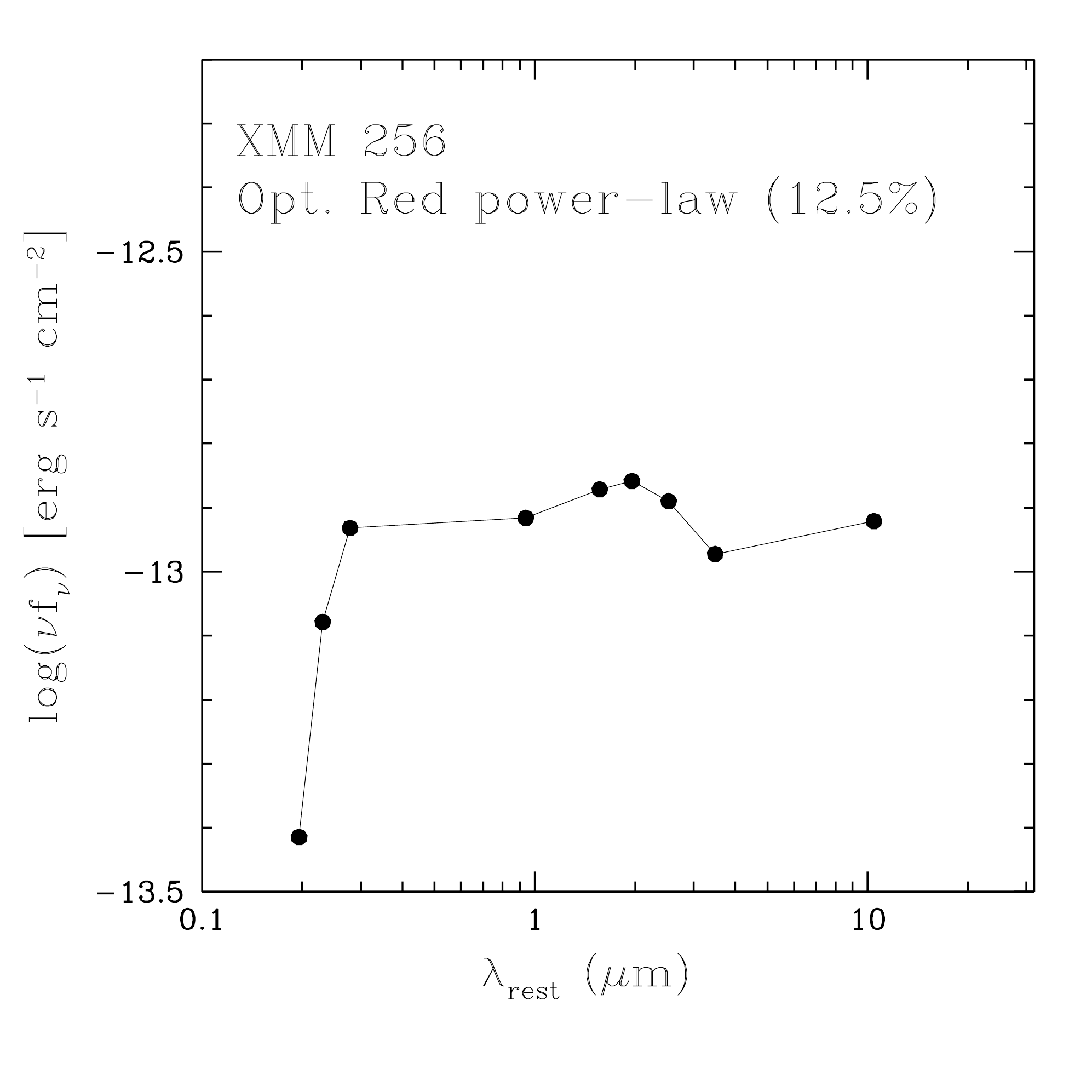}\\
\includegraphics[width=4.2cm]{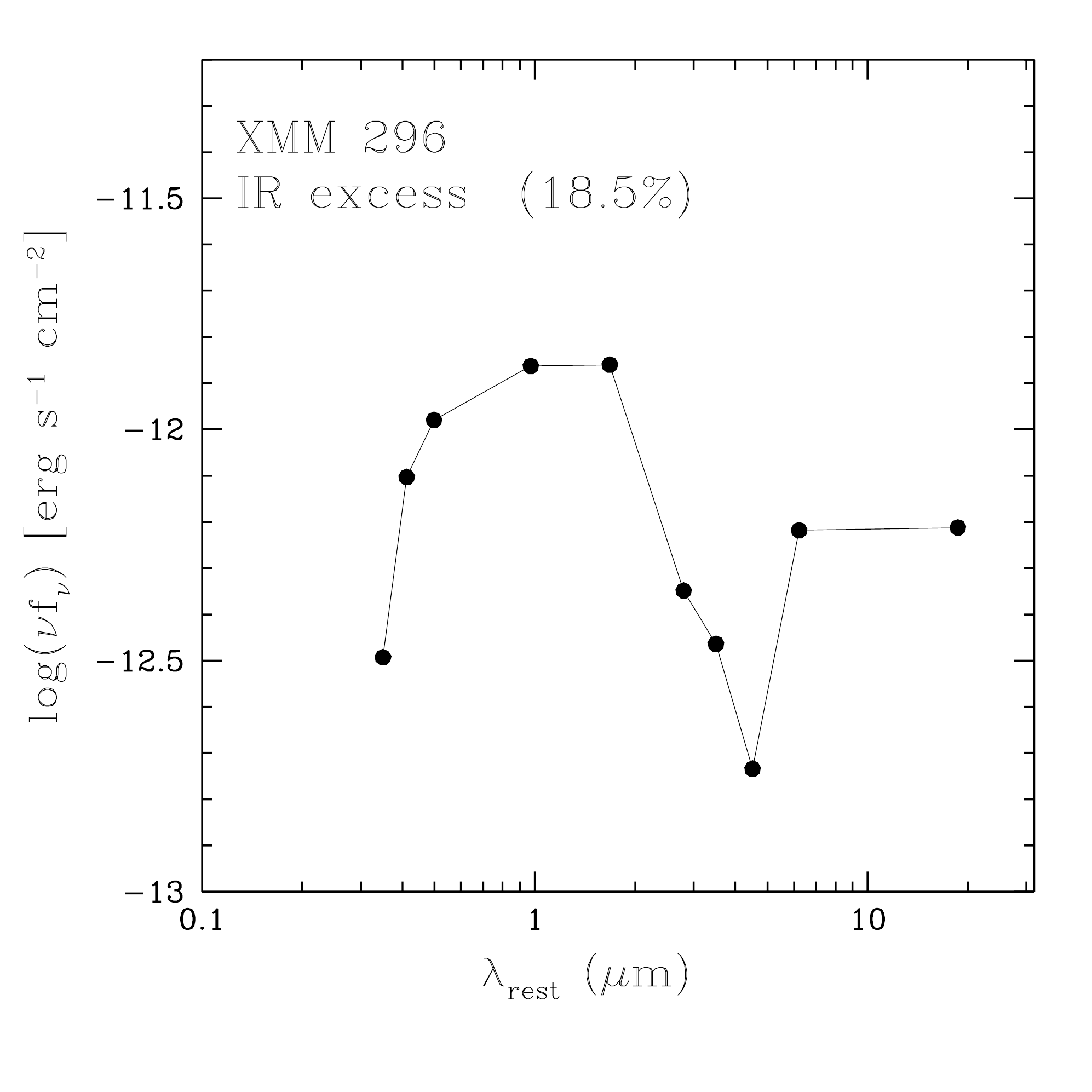}
\includegraphics[width=4.2cm]{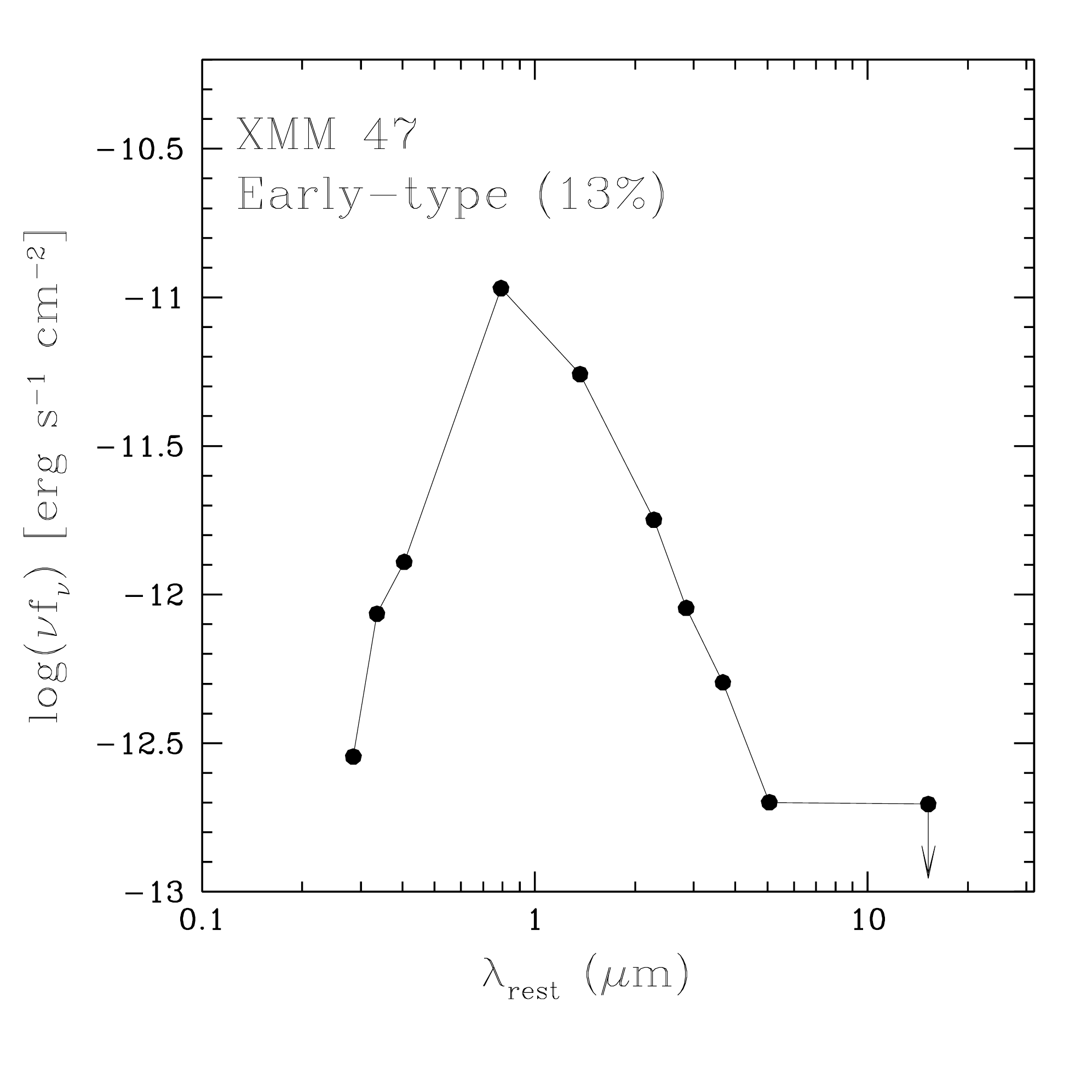}\\
\end{tabular}
\caption{Rest-frame spectral energy distributions for some X-ray sources. Upper limits are shown as arrows. Fractions are given respect to the whole X-ray sample.}
\label{essed}
\end{figure}

\subsection{Spectral energy distributions}\label{sed}
X-ray  sources show a variety of spectral energy distributions, which are not simple power-laws, typical of unobscured AGNs.
We empirically classified the X-ray sources SEDs in four classes:
 
\begin{itemize}

\item Optically blue power-law: objects whose SED resembles that of optically selected QSOs (the median QSO of \citet{Elvis1994}). These sources show a power-law SED  from the UV to mid-infrared wavelengths , without significant absorption in the optical bands.

\item Optically red power-law: sources showing signs of absorption in the optical bands, i.e. much steeper optical and UV continua compared to blue power-laws.

\item Early-type-like: the SED recalls that of early-type galaxies, and is thus dominated by the stellar continua, showing a peak at 1-3  \micron~ and then a decline up to 24 \micron.

\item IR excess : the SED peaks at  1-3 \micron~, then it declines but rises again at $\lambda \gtrsim $6 \micron;


\end{itemize}

Figure \ref{essed} shows some examples of these SED.
Power-law SEDs are clearly dominated by the AGN emission.
The mid-infrared emission of NOT-power-law SEDs  could be either dominated by an AGN or by star formation, while the optical-to-near-infrared emission is dominated by the host galaxy stellar continua.

We were able to classify the SED shape for 319 sources (70\perc~ of the sample, excluding stars and the X-ray extended sources).
For the remaining, the sources have too few bands over the detection limit and the SEDs cannot be characterized.
About 39\perc~ of the X-ray sources shows either an optically blue or a red power-law SED.
Optically red power-law SEDs are $\sim$12\perc~ of the sample.  
These objects have significant absorption in the optical bands,
compared to the optically blue power-law SEDs, and therefore they may be optically obscured AGNs.
13\perc~ of the SEDs are consistent with early-type galaxies but in many cases the limit at 24 \micron~ is too shallow to discriminate between these SEDs and those rising at 24 \micron.
For 18.5\perc~ of the sample the SEDs show a peak at 1-3 \micron, a decline and then a rise at $\lambda \gtrsim 6$ \micron.

The SED classification can be compared with the optical spectroscopic identification.
Power-law SEDs trace BL AGN. NOT BL AGNs show a variety of SED shapes, only 26\perc~ of them being power-laws.
The correspondence between SED and spectroscopic classification is summarized in Table \ref{tabsed}.
In the redshift interval z=0.8-1.5, sources with a power-law SED are the dominant population at high luminosity (\Lx$\ge 10^{44}$ \ergs), making up 82\perc~ of the sample.
Conversely, for \Lx$<10^{44}$ \ergs power-laws are only 47\perc~ of the total population.
The R magnitude and 2-10 keV flux distributions of the power-law 
sources without redshift and of the BL AGNs are different, the former being shifted towards fainter magnitudes (the Kolmogorov-Smirnov test probability that they are drawn from the same parent population is $<$0.1\perc). This is related to the magnitude selection of the spectroscopic targets. 
The same holds for the NOT power-law sources with and without redshift. 
This selection effect must be taken into account when
using the spectroscopic sample to compute luminosity functions or other evolution related quantities.

\begin{table}[b]
\caption{Number of BL/ NOT BL AGNs and total sources showing the different SED shapes. }\label{tabsed}
\begin{center}
{\tiny
\begin{tabular}{lcccc}
\hline
 &  Blue PL & Red PL & ET & IR-excess \\
\hline
BL AGN & 61 & 14 & 4 & 11 \\
NOT BL AGN & 15 & 15 & 27 & 38 \\
No Class & 43 & 28 & 28 & 35 \\
\hline
Total & 119 & 57 & 59 & 84 \\
\hline
\end{tabular}
}
{\tiny PL=power-law SED, ET=early-type-like.}
\end{center}
\end{table}

\begin{figure}
\centering
\begin{tabular}{c}
\includegraphics[width=8.5cm]{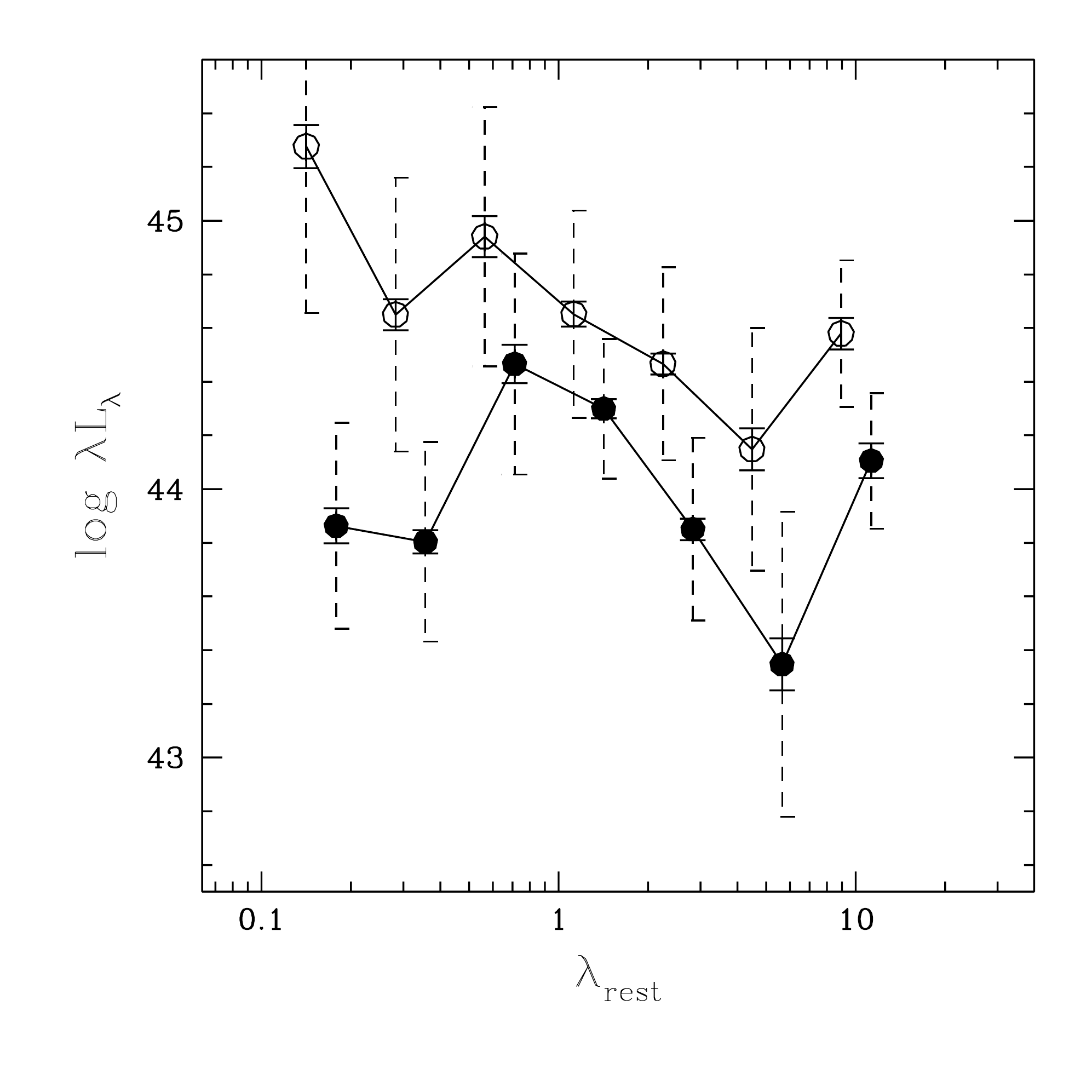}\\
\includegraphics[width=8.5cm]{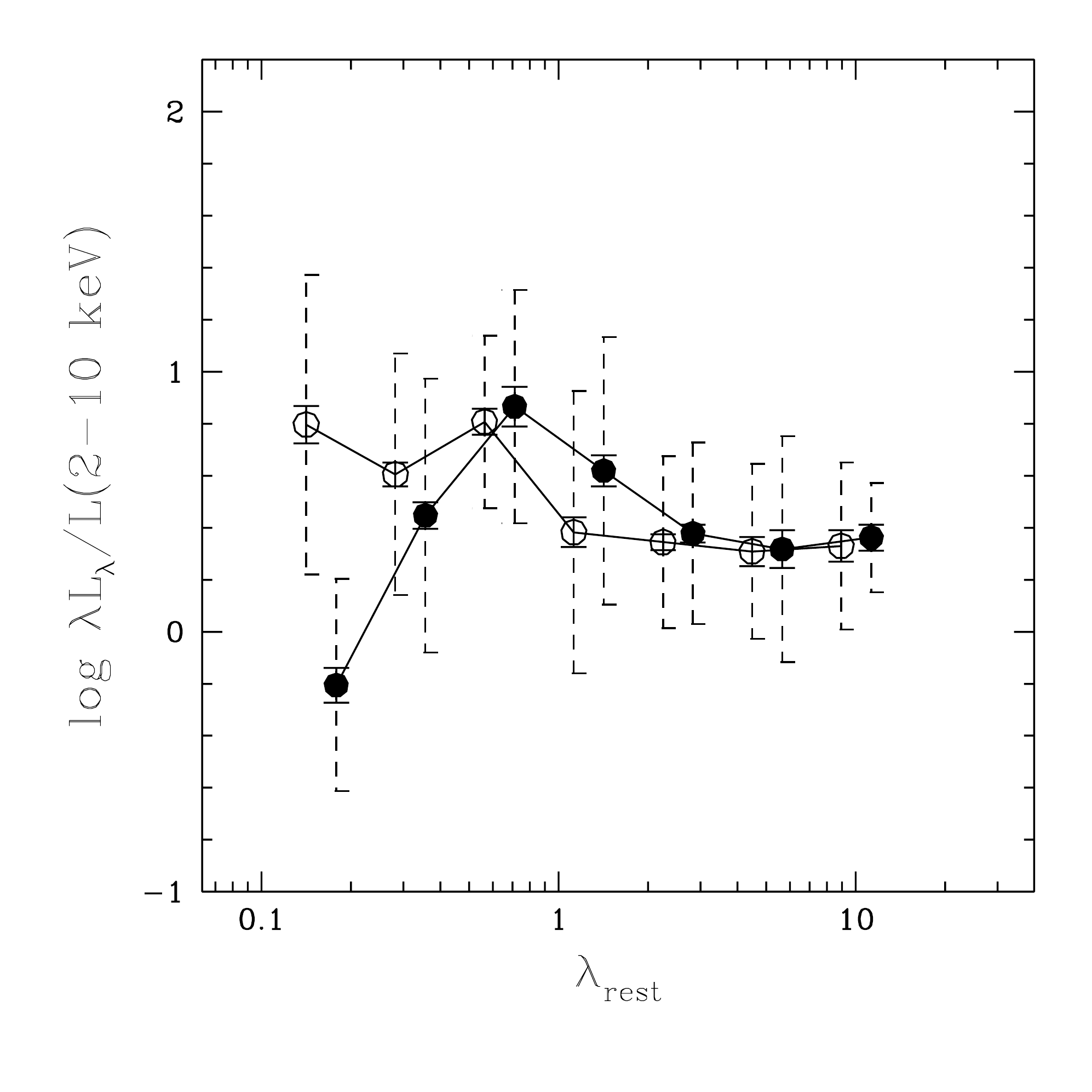}
\end{tabular}
\caption{ Average rest-frame SEDs for BL AGNs (open circles) and NOT BL sources (filled circles). The lower panel shows the average SEDs normalized to the 2-10 keV luminosity. }
\label{sedmed}
\end{figure}

\subsection{Average SEDs}
Figure \ref{sedmed}, upper panel, shows the average rest-frame SED for the sources spectroscopically classified as BL AGN and NOT BL AGN, for which we computed the intrinsic rest-frame monochromatic luminosity for each photometric data point available.
The average and dispersion were computed taking into account the upper limits on the fluxes in each band following \citet{Schmitt1985}.
The error on the mean and the dispersion are shown as dashed and solid error-bars, respectively.
The average SED of the BL AGN resembles a power-law, and the UV luminosity is consistent with the presence of an unobscured active nucleus. Indeed 3/4 of the BL AGNs have a power-law SED.
Conversely, NOT BL AGNs show an average SED dominated by the galaxy stellar light, with a peak at $\sim$1 \micron~ and a decline long-wards of 1-2 \micron, suggesting that the putative active nucleus is highly obscured. 
The rise of the continuum above 10 \micron~ suggests that the far-infrared emission starts to be dominated by the AGN emission.

Figure \ref{sedmed}, lower panel, shows that the average logarithmic 
ratio between the 10 \micron~ and the 2-10 keV luminosity is
about 0.4, with a dispersion of about half dex, for both BL and NOT BL AGNs.
Similar ratios are found for the Hellas2XMM highly obscured AGN of \citet{Pozzi2007}.

\subsection{Extended X-ray sources}\label{ext_sec}
Seven extended sources have been identified in the field by \citet{Puccetti2006}. 
For six of these sources we determined the cluster redshift from the optical spectrum of the brightest cluster galaxy. For the cluster \XMM 374 a photometric redshift is available.
Six out of seven of the optical counterparts have a SED dominated by the stellar emission, as expected from the cluster CD galaxies, and show a peak in the K band (six galaxies have K$<$16.2).
Five sources show an optical spectrum typical of early-type galaxies, without emission features.
Only \XMM 184 shows strong emission lines in its spectrum (H$\alpha$, H$\beta$ and [OIII]).
The results are summarized in Table \ref{clusters}. 
The spectroscopic redshifts agree with the photometric ones in \citet{Puccetti2006} within the errors, except for the source \XMM 363. 
For these two clusters we give a new estimate of the 0.5-10 keV luminosity based on the newly determined spectroscopic redshifts (see Table \ref{clusters}).
L(0.5-10 keV) is calculated using the total band fluxes and temperature values reported in \citet{Puccetti2006}, 
a thermal bremsstrahlung model without absorption, and the newly determined redshifts. 
Interestingly, four extended sources have redshift $\sim$0.39, which corresponds to
the highest peak in the spectroscopic redshift distribution of both X-ray (Figure \ref{istz}) and K-selected sources (Sacchi et al. 2008), suggesting the presence of a significant large scale overdensity.

\begin{table}[b]
\caption{ELAIS-S1 extended sources.}\label{clusters}
\begin{center}
{\tiny
\begin{tabular}{cccccc}
\hline
\hline
       Source name  &   R    &        K   &     F(0.5-10 keV)   &    z   &     logL$_X$    \\
                                 &  mag & mag    &      c.g.s                  &         &     erg s$^{-1}$\\
\hline
XMMES1\_145    &    16.9   &    13.7     &  1.46e-13     &   0.223        &     43.3  \\
XMMES1\_148    &    17.5   &    14.3     &    6e-15         &  0.278         &     42.1  \\
XMMES1\_184    &    20.6    &   18.4     &   8e-15          &  0.3877      &      42.6  \\
XMMES1\_224    &    17.8    &   14.2     &  1.43e-13      &  0.3925      &     43.9   \\
XMMES1\_363    &    20.2    &   16.2     &  1.6e-14        &  0.6338      &     43.4   \\
XMMES1\_374    &    19.3    &   16.0     &  1.2e-14        &      0.40$^{+0.41}_{-0.22}$*         &     42.8   \\
XMMES1\_444    &    18.3    &   14.7     &  2.5e-14        &     0.396      &     43.1   \\
\hline
\end{tabular}
}
{\tiny (* photometric redshift)}
\end{center}
\end{table}

\section{Multiwavelength  properties}

\begin{figure}[t]
\centering
\begin{tabular}{c}
\includegraphics[width=8.5cm]{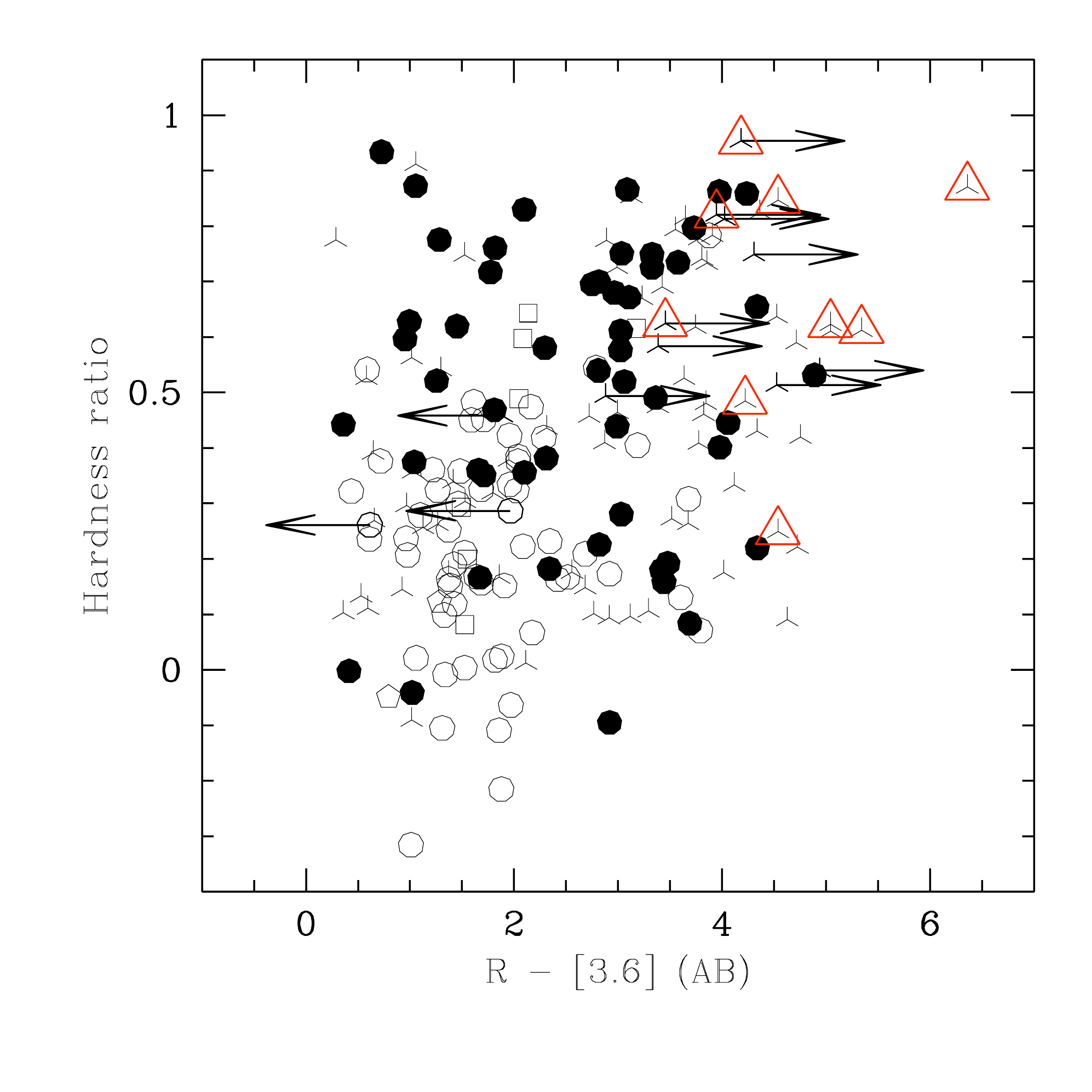}\\
\end{tabular}
\caption{Hardness ratio (HR=(Hard-Soft)/(Soft+Hard)) as a function of R-[3.6 \micron] color for BL AGN (open symbols) and NOT BL AGNs (filled symbols) from the 2-10 keV sample.
Sources without a spectroscopic redshift are shown as skeletons. Arrows = upper/lower limits; open triangles = sources with \miro$>1000$. 
}
\label{hrtrmk}
\end{figure}

\begin{figure}[t]
\centering
\begin{tabular}{c}
\includegraphics[width=8.5cm]{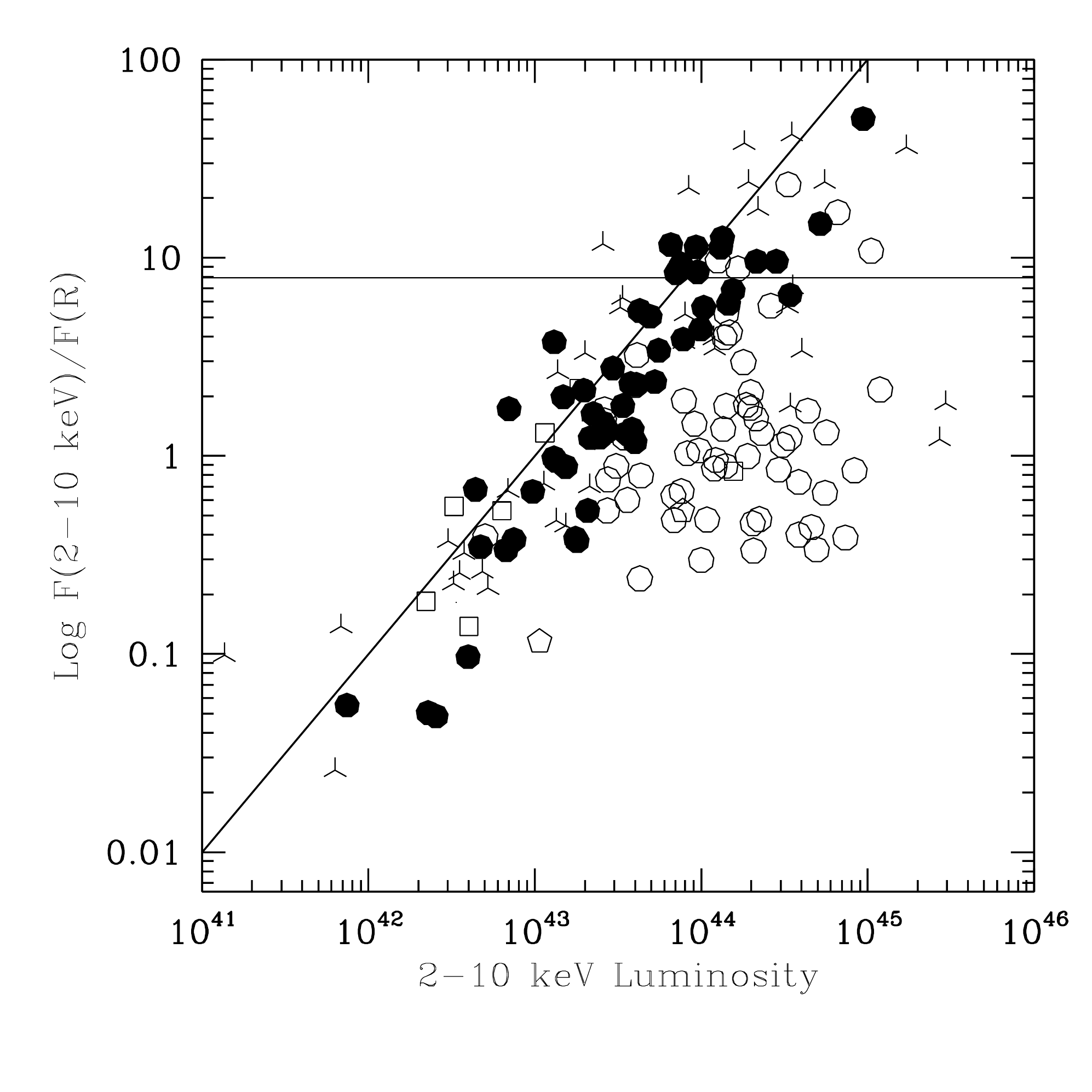}\\
\includegraphics[width=8.5cm]{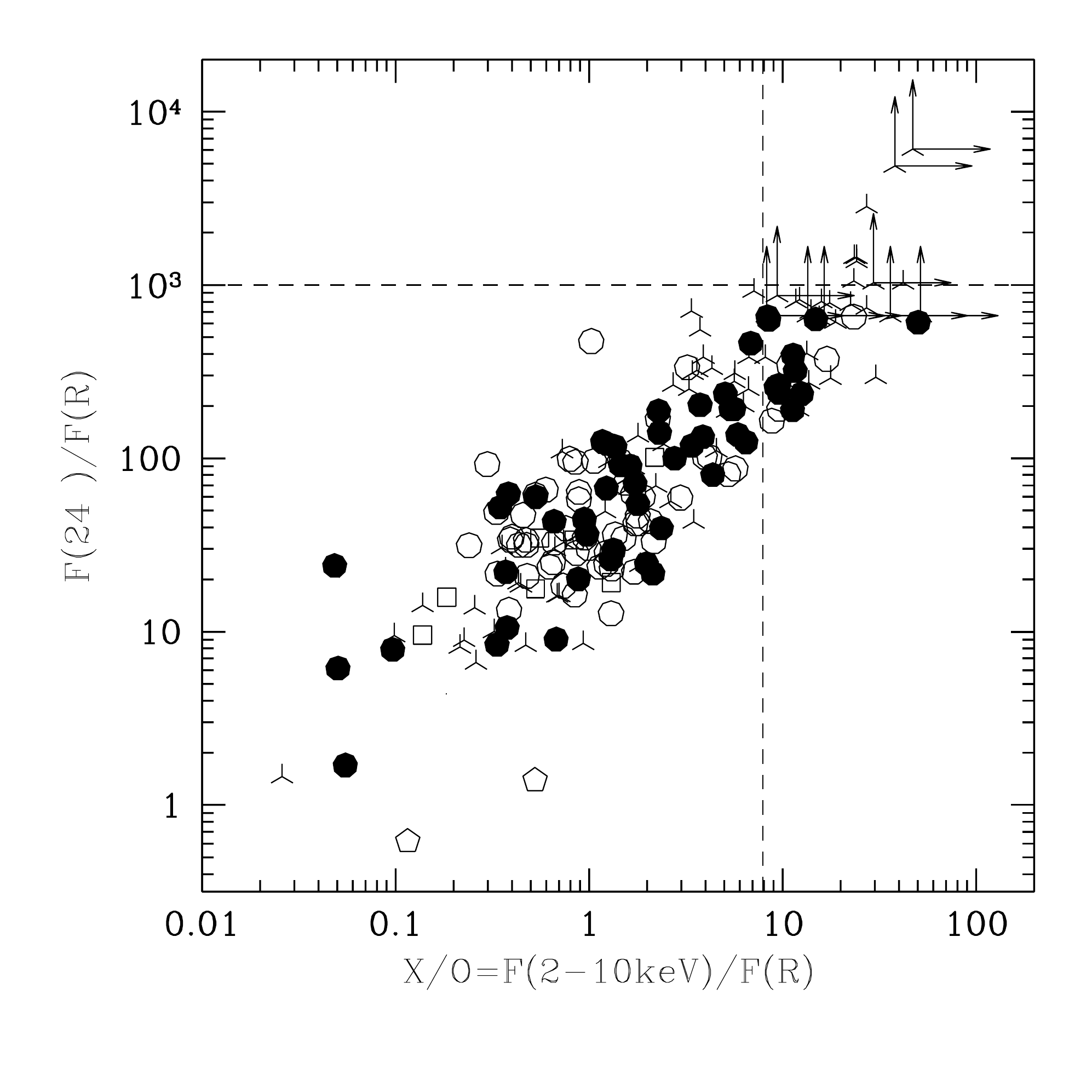}\\
\end{tabular}
\caption{Upper panel: the correlation between log$(f_X/f_R)$ and the 2-10 keV luminosity for type 2 objects (filled symbols). BL AGNs are shown as open symbols. The solid line  is a linear regression to the data of Fiore et al. (2003). Sources with a photometric redshift are shown as skeleton triangles.  Bottom panel: the correlation between X/O and the  mid-infrared-to-optical flux ratio (\miro).}
\label{xolx}
\end{figure}

\begin{figure}[h!]
\centering
\begin{tabular}{c}
\includegraphics[width=8.5cm]{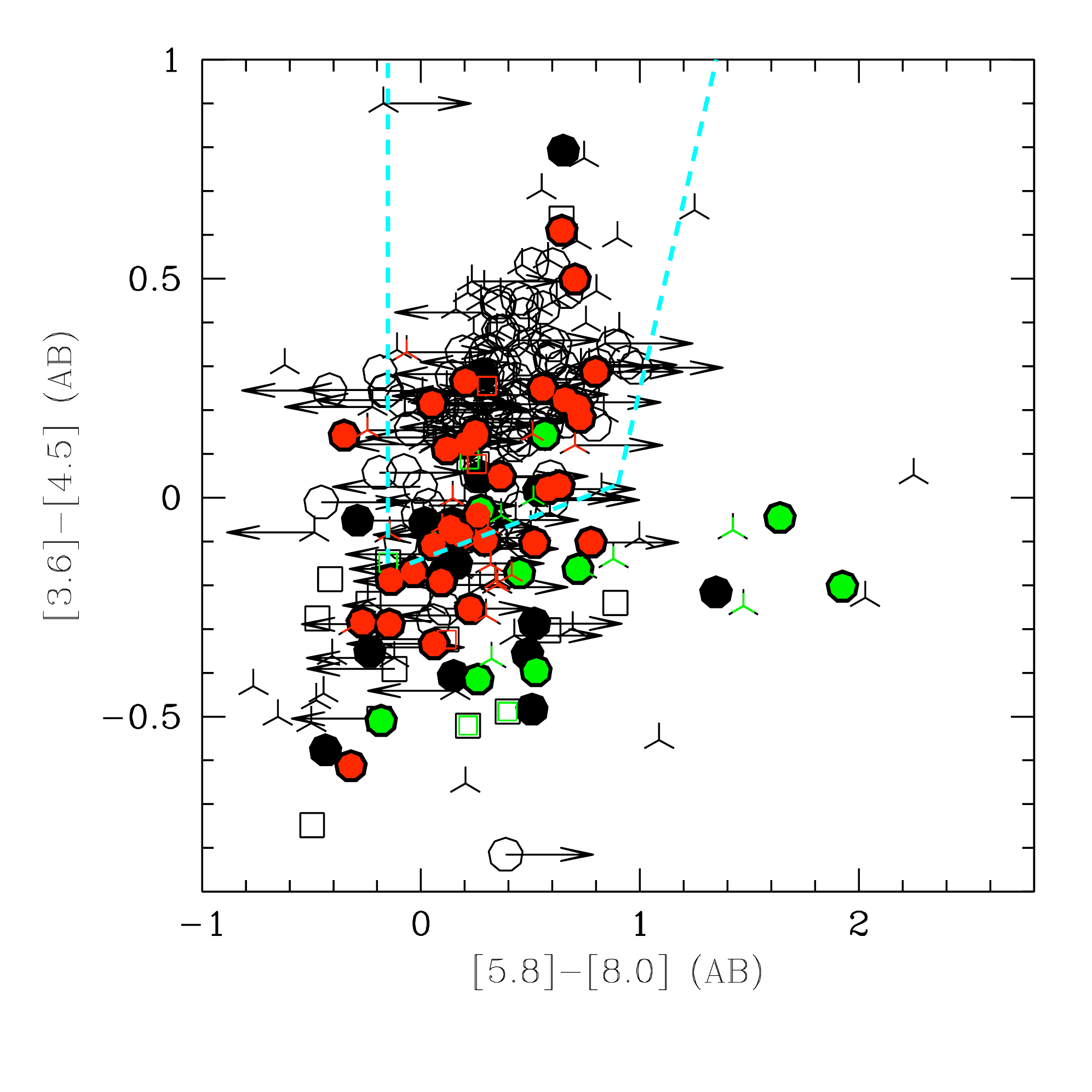}\\
\includegraphics[width=8.5cm]{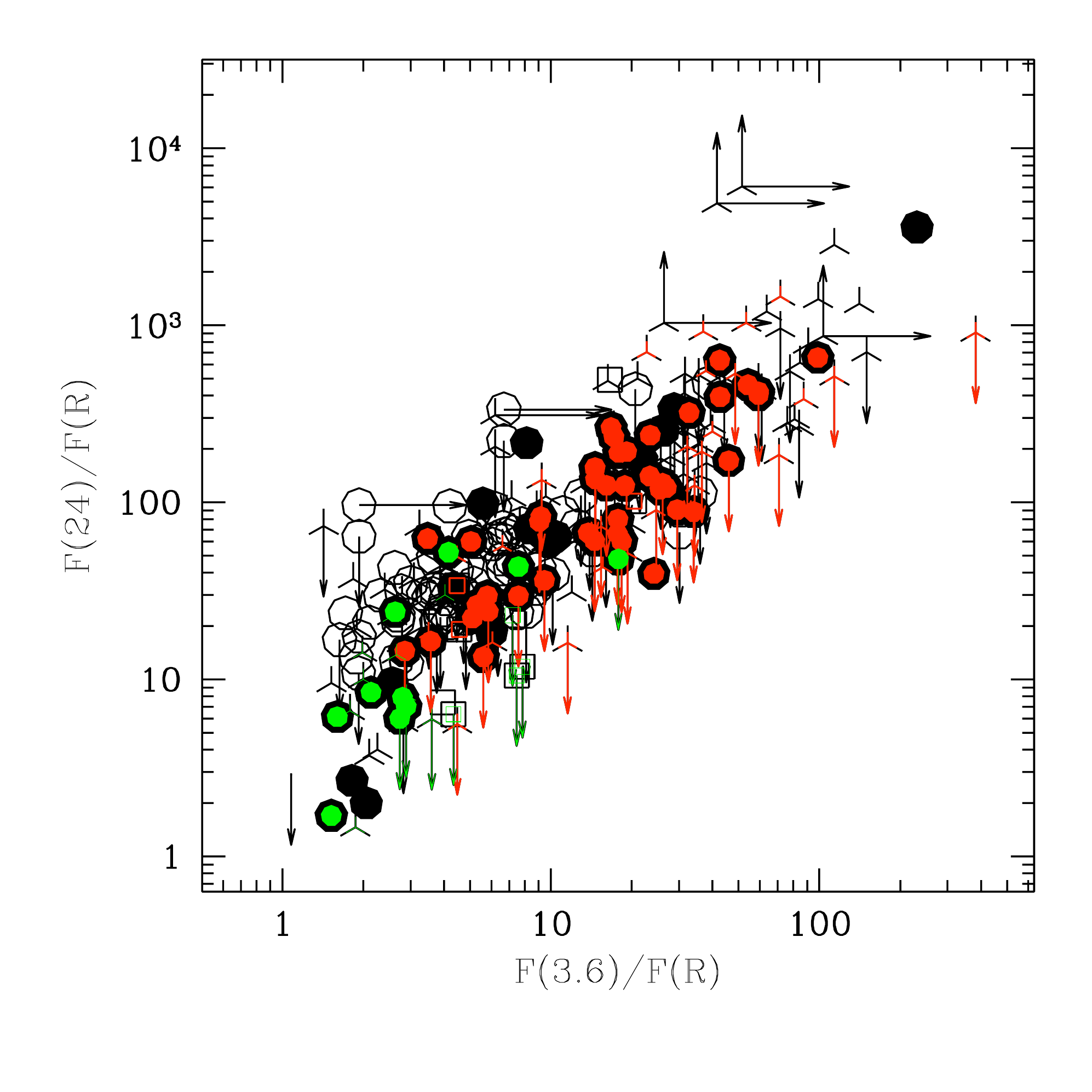}\\
\end{tabular}
\caption{Upper panel: IRAC color-color plot for the X-ray sample. 
The locus proposed by \citet{Stern2005} to select AGN is indicated by the dashed line. 
NOT BL low-luminosity ($41.8<$log\Lx$<43$ \ergs) AGNs are shown as green symbols, while high-luminosity NOT BL AGNs (log\Lx$>43$) are represented by red symbols.
Black filled symbols are NOT BL AGNs without a detection in the 2-10 keV band.
Open circles represent BL AGNs.
Sources with a photometric redshift are shown as skeleton triangles. 
Upper/lower limits are indicated by arrows.
Lower panel: the correlation between \miro and F(3.6 \micron)/F(R). 
Same symbols as in upper panel.}
\label{stern}
\end{figure}

Figure \ref{hrtrmk} shows the hardness ratio as a function of the $R-[3.6~\mu m]$ color of the sources drawn from the 2-10 keV sample. 
BL AGNs (open circles) are significantly softer than NOT BL AGNs (filled circles), with median $HR=0.23[0.12]$ vs. $HR=0.53[0.20]$, respectively, in agreement with \citet{Perola2004}.
The same figure shows that BL AGNs have, on average, bluer $R-[3.6~\mu m]$ colors compared to NOT BL AGNs, with median $R-[3.6~\mu m]$=1.71[0.41] for BL, and median $R-[3.6~\mu m]=3.0[1.1]$ for NOT BL AGNs, in agreement with \citet{Brusa2007}.
The reddest NOT BL sources are, on average, found at higher redshift with respect to bluer sources of the same class: median $z=1.71[0.36]$ for $R-[3.6\mu m]>3$, versus a median $z=0.46[0.22]$ for the NOT BL sources with bluer $R-[3.6~\mu m]$ colors.
In the same plot we also show the sources with \miro$>1000$ as open triangles.
These sources are among the reddest of the sample, and they have, on average, harder X-ray colors compared to those with \miro$<1000$. 
They show a median $HR=0.6[0.1]$),
which corresponds to an intrinsic absorbing column $log N_H=22.47^{+0.19}_{-0.24}~cm^{-2}$ at $z=1$. 
Four out of nine have $HR>0.8$, and three have $HR$ compatible with $HR=1$ within the errors ($log N_H=23.6~cm^{-2}$ at $z=1$).
Among the sources with $HR>0.8$, 27\perc~ have \miro$>1000$. 
Conversely $<3$\perc~ of the sources with $HR<0.8$ have \miro$>1000$.

Figure \ref{xolx}, upper panel, shows X/O versus the 2-10 keV luminosity for the 2-10 keV sources with either a spectroscopic or a photo-z. 
NL sources (filled symbols) follow the \citet{Fiore2003} correlation between X/O and \Lx. BL AGNs (open symbols) do not show any correlation. 
47 sources of the 2-10 keV sample have X/O$>$8.
Twenty of these have a redshift, either spectroscopic or photometric. 
Eleven out of 16 of the sources with a spectroscopic redshift are NOT BL AGNs, and all have 2-10 keV luminosity higher than 6.6$\times 10^{43}$ \ergs (ranging from log\Lx$=43.82$ to $\sim 45$ \cgs). 
Other four sources have a photometric redshift, and their X/O together with their SED makes them {\em bona fide} NL AGN.
The NL AGN with the highest luminosity is XMM 453 with $z=2.1733$ and  log\Lx$=44.97$ \cgs (see Figure \ref{spettri} for its optical spectrum). 
It has HR=0.67, corresponding to a rest frame N$_H\sim10^{23}$ cm$^{-2}$.
This source can therefore be considered an example of
type 2 QSO. 
Other similar objects are XMM 153 (z$=2.561$, log\Lx$=44.7$ and N$_H=3.9\times10^{23}~cm^{-2}$), XMM 101 (z$=1.772$, log\Lx$=44.4$ and N$_H=8.3\times10^{22}~cm^{-2}$) , XMM 121 (z$=1.645$, log\Lx$=44.3$ and N$_H=8.9\times10^{22}~cm^{-2}$).
The ELAIS-S1 sample confirms that high luminosity type 2 AGN can be efficiently selected among the sources with X/O$>8$.
However, 67\perc~ of the sources with X/O$>8$ remain spectroscopically unidentified  due to their faint optical magnitudes ($\sim$1/3 of them are fainter than $R=24.5$ and $\sim$2/3 have an average \R=24.2).
The sources with the highest X/O values show also the highest \miro ratios (Figure \ref{xolx}, lower panel), which is itself a proxy for obscuration.
Indeed, recent results indicate that Compton-thick AGN,  detected in only very small numbers even in the deepest hard X-ray surveys, are efficiently selected among sources with AGN luminosity in the mid-infrared, and faint optical or near-infrared emission (see for example \citet{Weedman2006}).
\citet{Fiore2008} found that the majority (70 to 90\perc) of the sources with extreme \miro ratio are likely to be highly obscured QSO.   
The sources that show the highest X/O and \miro values are XMM 63 and XMM 123. These sources have very faint optical counterparts and therefore we could not measure their redshift through optical spectroscopy nor SED fitting.
It is interesting to note that they are also among the hardest sources in the sample, with HR=0.95 and 0.82, respectively.

Figure \ref{stern} (top panel) presents an IRAC color-color plot \citep
{Stern2005}, where
we show the X-ray sources which have a detection in at least two IRAC bands. 
About 2/3 of the 0.5-10 keV sample without considering upper limits (1/3 of the sources with a detection in the four IRAC bands) lies outside of the AGN region (dashed line), which is populated by almost all BL AGN and 
NOT BL AGN with high-luminosity (\Lx$>10^{43}$ erg cm$^{-2}$ s$^{-1}$, red symbols).
Low-luminosity NOT BL AGNs ($10^{41.8}<$\Lx$<10^{43}$ erg cm$^{-2}$ s$^{-1}$, green symbols) occupy an oblique stripe in the lower part of the diagram, with a broad range of [5.8]-[8.0] color (0$\lesssim$[5.8]-[8.0]$\lesssim$2), reflecting a SED with a minimum within the IRAC range.
The sources  with [5.8]-[8.0]$>$1 are optically bright galaxies with redshift in the range 0.1-0.3, for the PAH emission features at 6.2 and 7.7 \micron~ are strong and redshifted in the 8.0 \micron~ IRAC band.
Their rest-frame U-V color is about 1.5, suggesting a moderate star formation activity. This is confirmed by their moderate F24/F(K) ratio.
These objects are therefore likely to be low-luminosity AGN in moderate star forming galaxies.
The low luminosity NOT BL AGNs with [5.8]-[8.0]$<$1 also have z$<$0.3 and they are hosted in somewhat less active galaxies (U-V$_{restframe}>$ 2).
In the region [5.8]-[8.0]$<$0 and [3.6]-[4.5]$<$0 we identified 5 early type galaxies at redshift between 0.28 and 0.72 with very little or no [OII] emission (XMM 84, 214, 430,149, 317). 
Their 0.5-2 keV X-ray luminosity ranges from 10$^{42}$ and 10$^{42.7}$ \ergs.
Four of these galaxies are covered by \chandra~ observations, and
all of them are detected. In three cases the emission is unresolved and centered on the galaxy nucleus. These could therefore be XBONG \citep{Comastri2002}.
In one case the emission is slightly elongated or due to two blended sources \citep{Civano2007}.

High luminosity NOT BL AGNs are found at a systematically higher redshift
(85\perc~ in the range z$=$0.6-1.4), 
At these redshifts the V rest frame is shifted within the observed J band.
Unfortunately, J magnitudes are available only for a small fraction of
the sources due to an incomplete coverage of the field. Therefore
we could derive U-V rest frame colors for only half of the high-luminosity NOT BL AGNs. 
Nevertheless, they indicate a moderate star forming galaxy also in these cases.   
To further investigate the galaxy colors of this source sample, we plot in 
Figure \ref{stern}, bottom panel, the correlation between \miro and \niro.
High luminosity sources span a wide range of \miro and \niro, reaching values close to \miro=10$^3$. 
Note that both colors are systematically higher than those of low-luminosity AGN. 
The high \miro can be due to AGN emission dominating the 24 \micron~ band.
The quality of our photometry is not good enough to understand whether the red [3.6]-R color is due to a passive host or to a dusty star-forming galaxy.

\section{Summary}
We have presented the optical and infrared identifications of 478 X-ray sources detected by \xmm~ in the central 0.6 deg$^2$ of the ELAIS-S1 field.
The identification process was validated by precise (arcsec) source positions obtained with \chandra~ in a fraction of the area covered by the \xmm~ survey. 
We found  that \chandra~ observations are crucial to identify the correct counterpart for optically faint sources. 
Comparing the \chandra~ identifications with the IRAC ones, we find that the latter miss only 4\perc~ of the real counterparts.
Therefore, we feel confident to use the IRAC identifications for the area not covered by \chandra.

We compiled a 
multiwavelength catalogue, with photometric data ranging from the mid-infrared to the optical bands. 
We obtained optical spectra to measure redshifts and to obtain a first classification of the counterparts.
The spectroscopy was performed using VIMOS/VLT, FORS2/VLT and EFOSC/ESO3.6m.
We obtained reliable 
redshifts and classification for 237 sources with optical counterparts brighter than  R$=$24.
47\perc~ of the sample turned out to be broad-line AGNs,
while the other sources are narrow-line AGNs (12\perc), ELGs (27\perc) and absorption-line galaxies (8\perc).
We find 47 sources showing \xo$>$8 (23\perc~ of the 2-10 keV sample) . 
Out of the 16 spectroscopically identified sources with \xo$>$8, 11 ($\sim$70\perc)
turned out to be type 2 QSOs at z=[0.9-2.6], with log\Lx$\ge$43.8 \ergs.
All these type 2 QSO have hard X-ray colors, suggesting large absorbing columns at the rest frame.
The ELAIS-S1 sample therefore confirms that type 2 QSOs are efficiently selected among the sources with high \xo.

We classified  empirically the broad band SEDs of the X-ray sources, from pure power-laws to galaxy-dominated SEDs. 
We find a generally good agreement between the SED classification and the optical spectroscopy. 
80\perc~ of the BL AGNs have power-law SEDs, and  $\sim$71\perc~ of the sources with  a power-law SED and a spectroscopic redshift are classified as BL AGNs.
 
Broad band SEDs have also been used to compute photometric redshifts. 
Reliable photometric redshifts  were obtained for a sample of 68 sources without a spectroscopic redshift and which show a SED dominated by the host galaxy stellar light.

We computed average rest-frame SEDs, finding that BL and NOT BL AGNs show similar 
L$_{10 \mu m}$/L(2-10 keV) ratios ($\sim$0.4 in logarithm), and consistent with \citet{Pozzi2007}.

By comparing near-infrared colors, we find that
low luminosity NOT BL AGNs (log\Lx$<$43 \ergs) are hosted in star forming galaxies, which show bluer rest-frame U-V and R-[3.6 \micron] colors.
High-luminosity NOT BL AGNs hosts (log\Lx$>$43 \ergs) have, on average, redder R-[3.6 \micron] colors. This colors can be due either to a dusty star-forming host galaxy, or to a passive early-type host (\citet{Pozzi2007}, \citet{Mignoli2004}), but the quality of our photometry does not allow us to distinguish between the two. 
Deeper multiwavelength surveys, such as COSMOS and GOODS, are needed to asses the nature of these objects.

\begin{acknowledgements}
We are grateful to Bianca Garilli and Marco Scodeggio for the support provided in running the VIPGI pipeline, and to Mary Polletta for providing template SED. 
This work is based on observations made with the Spitzer Space Telescope,
which is operated by the Jet Propulsion Laboratory, California Institute of Technology under a contract with NASA.
We aknowledge financial contribution from contract ASI-INAF I/023/05/0, PRIN-MUR grant 2006-02-5203 and the ANR grant D-SIGALE ANR-06-BLAN-0170.
\end{acknowledgements}

\begin{appendix}
\section{Multiband catalog}
Table \ref{cata} summarizes the entries of the catalogue of the 478 X-ray sources with multiwavelength data in 10 photometric bands, from B to MIPS 24 \micron~.
The complete catalogue is available in ascii format in the online version, or can be accessed  at the following URL:  http://www.oa-roma.inaf.it/ELAIS-S1/,
along with image thumbnails of the counterparts in four bands (R, K, IRAC 3.6 \micron~ and MIPS 24 \micron) and the optical spectrum where available.
Columns 1 to 11 present the X-ray data of the source: the \xmm~ source name, the $\alpha$ and $\delta$ (J2000) XMM positions, the 0.5-10, 2-10 and 0.5-2 keV fluxes and the respective S/N ratios as in the \citet{Puccetti2006} catalogue, and the \chandra~ positions. 
Columns from 12 to 18 give the optical and near-infrared J and K band photometric data (R, J, K Vega magnitudes, magnitudes below the detection limit are negative).
Columns  20 to 28 give the coordinates and 
flux densities in the four IRAC channels and MIPS 24 \micron~ channel, in $\mu$Jy).
Columns 29 to 33 report the SED, source redshift and classification.
Figure \ref{webpage} shows an example of the online pages.

\begin{figure}[t]
\centering
\begin{tabular}{c}
\includegraphics[width=8.5cm]{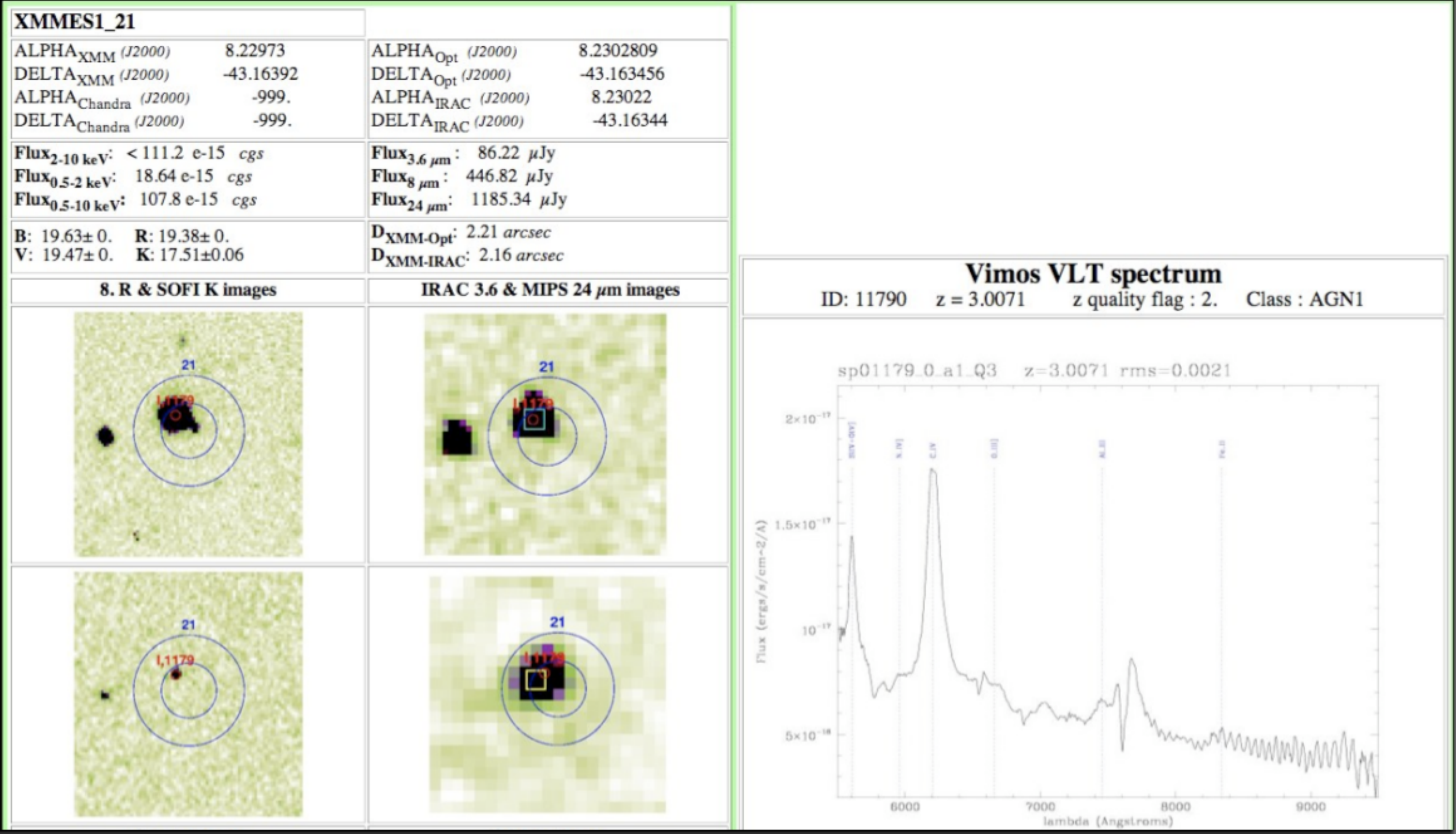}\\
\end{tabular}
\caption{ The appearance of the online catalogue pages. Source images in four bands (R, K, IRAC  3.6 \micron~ and MIPS 24 \micron~) and the optical spectrum are shown, together with the source position, fluxes and magnitudes.}
\label{webpage}
\end{figure}

\begin{table}
\caption{Data fields available in the online multi-band catalogue.}
\begin{center}
{\tiny
\begin{tabular}{|rrl|}
\hline
\hline
1 & NAME & XMM source name \\
2 &RA\_X & XMM Right Ascension (J2000)\\
3 &DEC\_X & XMM Declination (J2000)\\
4 &F(0.5-10 keV) & 0.5-10 keV Flux [cgs/1e-15]\\
5 &S/N &  S/N Ratio at 0.5-10 keV \\
6 &F(2-10 keV) & 2-10 keV Flux [cgs/1e-15]\\
7 &S/N & S/N Ratio at 2-10 keV \\
8 &F(0.5-2 keV) & 0.5-2 keV Flux [cgs/1e-15]\\
9 &S/N & S/N Ratio AT 0.5-2 keV \\
10&RA\_Xch &  \chandra~ Right Ascension (J2000)\\
11&DEC\_Xch & \chandra~ Declination (J2000)\\
12&RA\_R &  R band Right Ascension (J2000)\\
13&DEC\_R & R band Declination (J2000)\\
14&R & R magnitude [Vega mag]\\
15&PROB & R band chance coincidence prob.\\
16&DD\_OPT & XMM-Optical displacement [arcsec]\\
17&J & J band magnitude [Vega mag]\\
18&K & K band magnitude [Vega mag]\\
20&RA\_IRAC & IRAC Right Ascension (J2000)\\
21&DEC\_IRAC & IRAC Declination (J2000)\\
22&FLUX\_36 & IRAC 3.6 \micron~ flux [$\mu$Jy]\\
23&FLUX\_45 & IRAC 4.5 \micron~ flux  [$\mu$Jy]\\
24&FLUX\_58 & IRAC 5.8 \micron~ flux  [$\mu$Jy]\\
25&FLUX\_80 & IRAC 8.0 \micron~ flux  [$\mu$Jy]\\
26&FLUX\_24 & MIPS 24 \micron~ flux [$\mu$Jy]\\
27&DD\_IR & XMM-IRAC displacement [arcsec]\\
28&PROB IR & 3.6 \micron chance coincidence prob.\\
29&SED & Spectral Energy distribution\\
30&Z & Spectroscopic redshift\\
31&QUAL & redshift quality flag\\
32&CLASS & source classification\\
33&ZPHOT & Photo-z\\
\hline
\end{tabular}
}
\end{center}
\label{cata}
\end{table}

\end{appendix}

\end{document}